\documentclass[twocolumn,amsmath,amssymb,showpacs,nofootinbib]{revtex4-1}

\usepackage[usenames,dvips]{color}
\usepackage{graphicx}
\usepackage{dcolumn}
\usepackage{bm}

\newcommand{\bea}{\begin{eqnarray}}
\newcommand{\eea}{\end{eqnarray}}
\newcommand{\bt}{\textbf}

\newcommand{\phd}{\phantom{\dag}}
\newcommand{\ph}{\phantom{.}}
\newcommand{\up}{^{\phd}}
\newcommand{\noi}{\noindent}
\newcommand{\no}{\nonumber}

\begin{document}
\def\v#1{{\bf #1}}

\title{Topological superconductivity in Rashba semiconductors without a Zeeman field}
\author{Panagiotis Kotetes}
\email{panagiotis.kotetes@kit.edu}
\affiliation{Institut f\"{u}r Theoretische Festk\"{o}rperphysik and DFG-Center for Functional Nanostructures (CFN), Karlsruhe Institute of Technology, 76128 Karlsruhe,
Germany}

\vskip 1cm
\begin{abstract}
In this manuscript I present new hybrid devices based on multi-wire/channel Rashba semiconductors, which harbor Majorana fermions (MFs) \textit{without} a Zeeman field. In
contrast, magnetic fluxes, supercurrents or electric fields can be employed, yielding an enhanced device manipulability. The generic topological phase diagram for
two-nanowire/channel systems exhibits features of quantum criticality and a rich interplay of phases with 0, 1 or 2 MFs per edge. The most prominent and experimentally
feasible implementation relies on the already existing platforms of InAs-2DEG on top of a Josephson junction. Appropriate design of the latter device allows phases with 1
or 2 MFs, both detectable in zero-bias anomaly peaks with a single or double unit of conductance. 
\end{abstract}


\maketitle

\section{Introduction} 

The perspective of to\-po\-lo\-gi\-cal quantum compu\-ting (TQC) \cite{TQC} has motivated a plethora of pro\-po\-sals for enginee\-ring topological superconductors (TSCs),
mostly relying on semiconductors with strong Rashba spin orbit coupling (SOC) \cite{Semi,NW,multiband,Tewari and Sau,Doublewire,Klinovaja,Frustaglia,KotetesClassi}. Among
them, a device invol\-ving a Rashba nanowire (NW) \cite{NW} lies in the spotlight of current research. The latter setup requires a sufficiently strong Zeeman field in order
to enter the TSC phase with 1MF per edge. The first en\-cou\-ra\-ging zero bias anomaly (ZBA) MF-fin\-dings have been already re\-por\-ted \cite{Mourik,MFexperiments}, that
however, remain under intense debate \cite{MFexperiments2}. 

A promising route for resolving this controversy is to explore alternative TSC platforms which build upon the \textit{same} mate\-rials used in these expe\-ri\-ments but with
the Zeeman field replaced by a supercurrent flow, an electric field or, as shown recently, by a SOC with time dependent orientation \cite{Frustaglia}. The latter ingredients
appear less harmful for the bulk SC and possibly more versatile for TQC. For instance, existing TQC protocols rely on sufficiently strong antiparallel magnetic fields on a
nanoscale level \cite{Flensberg}, which can be difficult to achieve in the lab. Instead, harboring MFs \textit{all-electrically} can be advantageous for braiding and
developing TSC circuits. 

\begin{figure}[t]
\centering
\includegraphics[width=0.56\columnwidth]{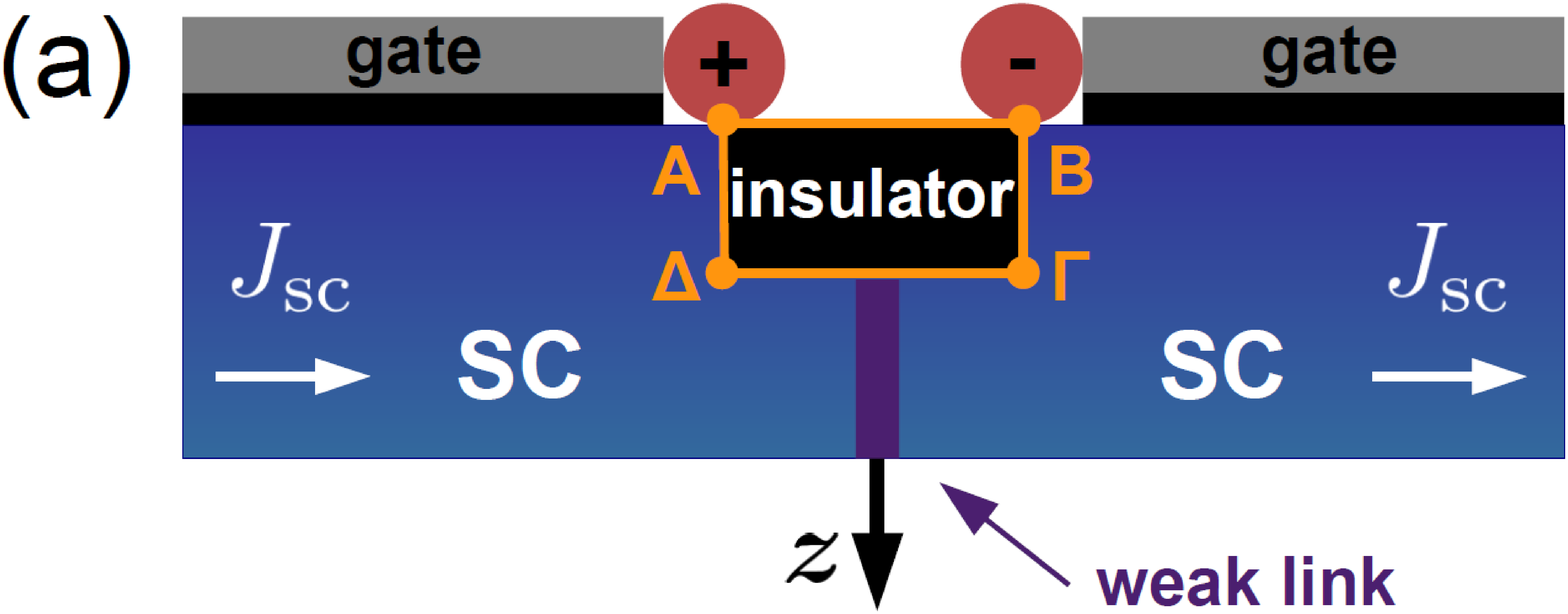}\hspace{0.25in}
\includegraphics[width=0.31\columnwidth]{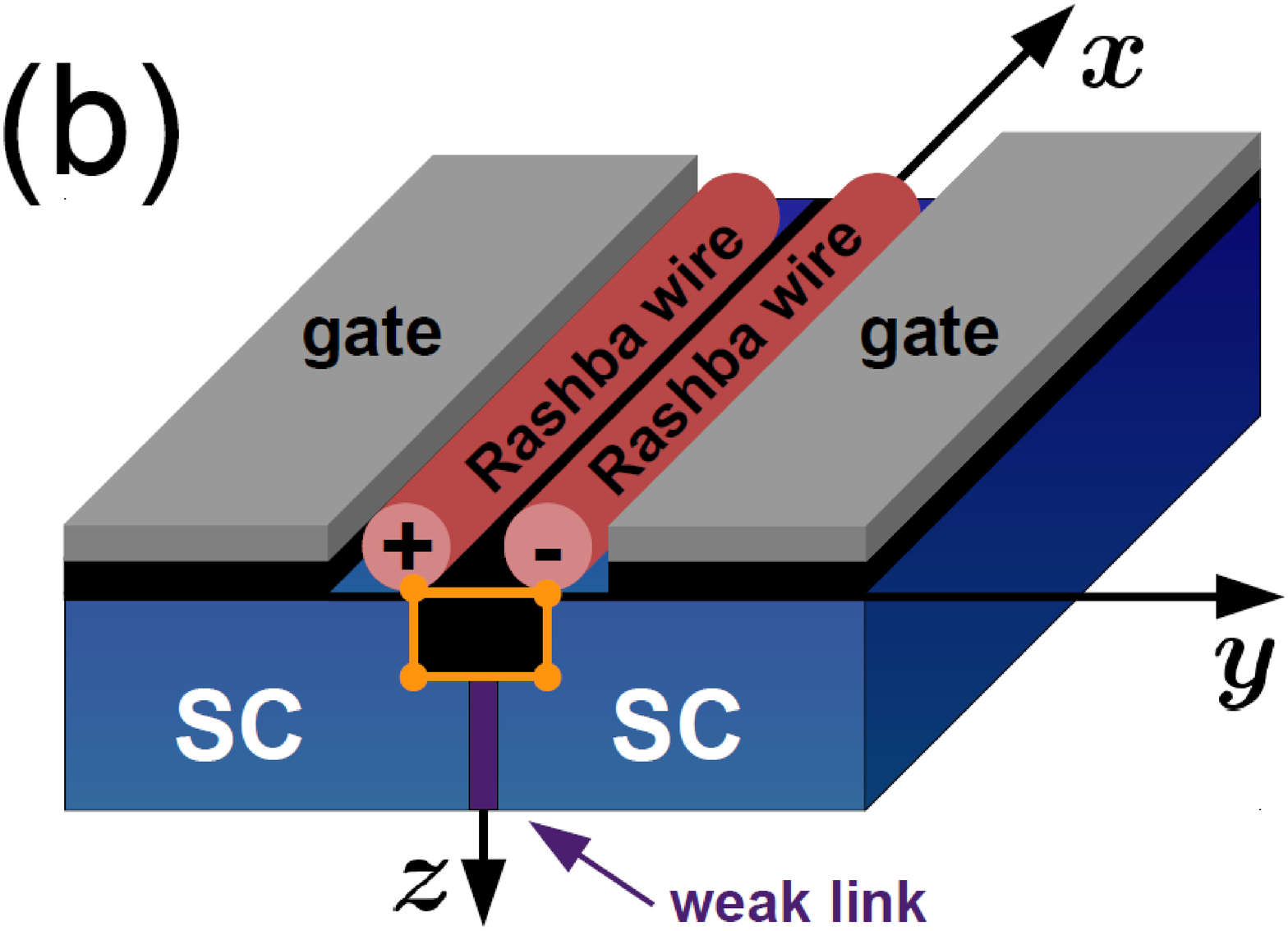}\\
\vspace{0.1in}
\includegraphics[width=0.72\columnwidth]{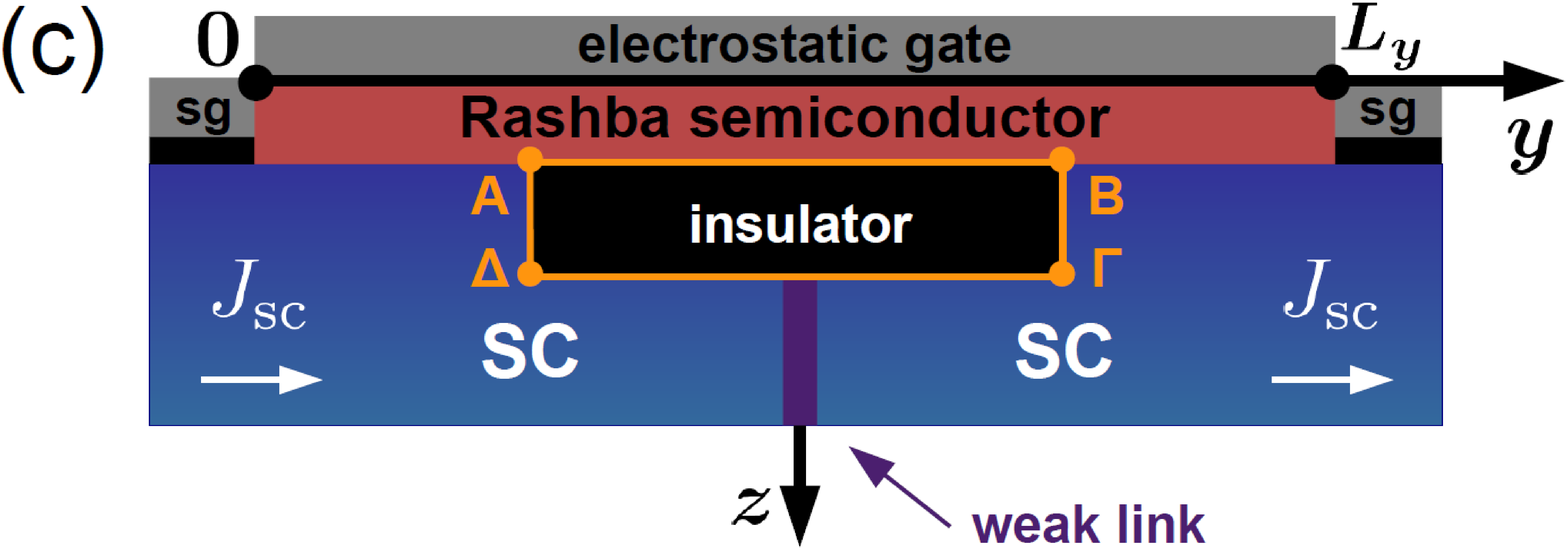}
\caption{(a/b) Side/Full view of two Rashba NWs and (c) Side view of a Rashba semiconducting film, on top of a Josephson junction. The interface is interrupted by an
insulating loop (AB${\rm \Gamma\Delta}$). Threading magnetic flux ($\Phi_{\rm flux}$) through the loop, leads to Majorana bound states extended along the $y$ axis and
localized at the edges along the $x$ axis. The required magnetic flux can be generated by: \bt{i.} a supercurrent flow ($J_{\rm sc}=2e\Phi_{\rm flux}/\hbar$) through the
junction or \bt{ii.} by placing the semiconductor in an electric field (${\cal E}_{AB}=\dot{\Phi}_{\rm flux}$), produced here by side gates (sg).}
\label{fig:Setups}
\end{figure}

In this work, I propose a new type of artificial TSCs consisting of conventional SCs in proximity to \textit{quasi-1d} semiconductors, such as InAs and InSb. The quasi-1d
aspect is instrumental for engineering a TSC \textit{without} a Zeeman field \cite{KotetesClassi}. The semiconductor can consist of either two coupled single-channel NWs 
or a multi-channel film/NW placed on top of a Josephson junction of two conventional SCs. Threading magnetic flux ($\Phi_{\rm flux}$) through the insulating loop
(AB${\rm\Gamma\Delta}$) depicted in Fig.~\ref{fig:Setups}, can lead to MFs. Apart from using a solenoid, one can generate the required flux via: \bt{i.} a supercurrent flow
$J_{\rm sc}$ through the Josephson junction or \bt{ii.} by placing the semiconductor in an electric field ${\cal E}_{AB}$. For the latter two implementations, the area of the
loop can be infinitessimaly small. Remar\-kably, functional devices similar to the one in Fig.~\ref{fig:Setups}(c) have been already realized with InAs two-dimensional
electron gas (2DEG) \cite{2DEGs}, rendering the present proposal highly feasible. 

The models of the two-NW/channel implementations which I consider here, can be mapped to each other and are characterized by a phase diagram suppor\-ting 0, 1 or 2 MFs per
edge. The phase with 2MFs per edge is protected by chiral symmetry \cite{Tewari and Sau,BDI} and meets the zero- and single-MF phases at two quantum tricritical points of 
the parameter space. By adopting parameter values representative of InAs and InSb, I make firm predictions for experimentally realizing and optimizing the present MF
platforms. I find that two coupled NWs can harbor a single MF per edge \textit{only} if they are in contact, while the most pro\-mi\-sing device relies on a two-channel
semiconducting film which supports both 1 and 2 MF-phases. The latter analysis can be also extended to a three-dimensional Rashba NW in the case where only the lowest three
channels become relevant and only two of them can exhibit non-trivial topological behavior. In fact, as I show here, the superconducting proximity effect has to be
substantially enhanced for enabling the latter devices to harbor MFs.

For pedagogical reasons, I first present in Sec.~\ref{Sec:NWs} the analysis of a hybrid device relying on two coupled single-channel NWs. In Sec.~\ref{Sec:2Cs} I continue
with the case of a semiconducting film with only two channels under consi\-de\-ration, which appears as the most experimentally feasible platform of the genre. In
Sec.~\ref{Sec:3Cs}, I extend my results to a three-dimensional Rashba NW when only three channels become relevant, and uncover the conditions which can allow the realization
of such TSC setups in the lab. Finally, I present my conclusions in Sec.~\ref{Sec:Conclusions}.

\section{Two coupled Rashba nanowires on top of a Josephson junction}\label{Sec:NWs}

In this section I first introduce (Sec.~\ref{SubSec:HD2NW}) the accurate three-dimensional Hamiltonian description of the device depicted in Figs.~\ref{fig:Setups}(a,b) 
and highlight (Sec.~\ref{SubSec:SuperFlux}) the intimate connection of threading flux through the loop (AB${\rm\Gamma\Delta}$) and imposing a supercurrent flow through the
Josephson junction. Later on, I retrieve an effective model for the coupled NWs that incorporates the pro\-xi\-mi\-ty induced superconducting gap
(Sec.~\ref{SubSec:EffectiveNWs}). For the latter model I perform a symmetry analysis (Sec.~\ref{SubSec:Symmetry}) which allows un\-co\-ve\-ring the relevant TSC mechanism
(Sec.~\ref{SubSec:TSCmech}). I conclude this section with extrac\-ting the detailed topological phase diagram (Sec.~\ref{SubSec:TPD}) and putting forward concrete predictions
for future experiments (Sec.~\ref{SubSec:ExpNWs}).

\subsection{Hybrid-device-Hamiltonian}\label{SubSec:HD2NW}

The model Hamiltonian corresponding to Fig.~\ref{fig:Setups}(a) reads:
${\cal H}=\int{\rm d}x \left[{\cal H}_{\psi}(x)+{\cal H}_{c}(x)+{\cal H}_{\psi c}(x)\right]$ with:
\bea
{\cal H}_{\psi}(x)&=&\sum_{n=\pm}\hat{\psi}_n^{\dag}(x)\left[\varepsilon(\hat{p}_x)+v\hat{p}_x\sigma_y\right]\hat{\psi}_n\up(x)
\no\\
&+&\big[\hat{\psi}_+^{\dag}(x)\left(t_{\perp}+iv_{\perp}\sigma_x\right)e^{-i\pi\phi}\hat{\psi}_-\up(x)+{\rm H.c.}\big]\,,\\
{\cal H}_{c}(x)&=&\sum_{n=\pm}\hat{c}_n^{\dag}(x)\tilde{\varepsilon}(\hat{p}_x)\hat{c}_n\up(x)+
\tilde{t}_{\perp}\big[\hat{c}_+^{\dag}(x)\hat{c}_-\up(x)+{\rm H.c.}\big]\no\\
&+&\sum_{n=\pm}\tilde{\Delta}\left[e^{niJ_{\rm sc}/2}c_{n\uparrow}^{\dag}(x)c_{n\downarrow}^{\dag}(x)+{\rm H.c.}\right]\,,\label{eq:HC}\\
{\cal H}_{\psi c}(x)&=&{\rm T}\sum_{n=\pm}\big[\hat{\psi}_n^{\dag}(x)\hat{c}_n\up(x)+\hat{c}_n^{\dag}(x)\hat{\psi}_n\up(x)\big]\,.\label{eq:HPsiC}
\eea

\begin{figure}[t]
\centering
\includegraphics[width=0.4\columnwidth]{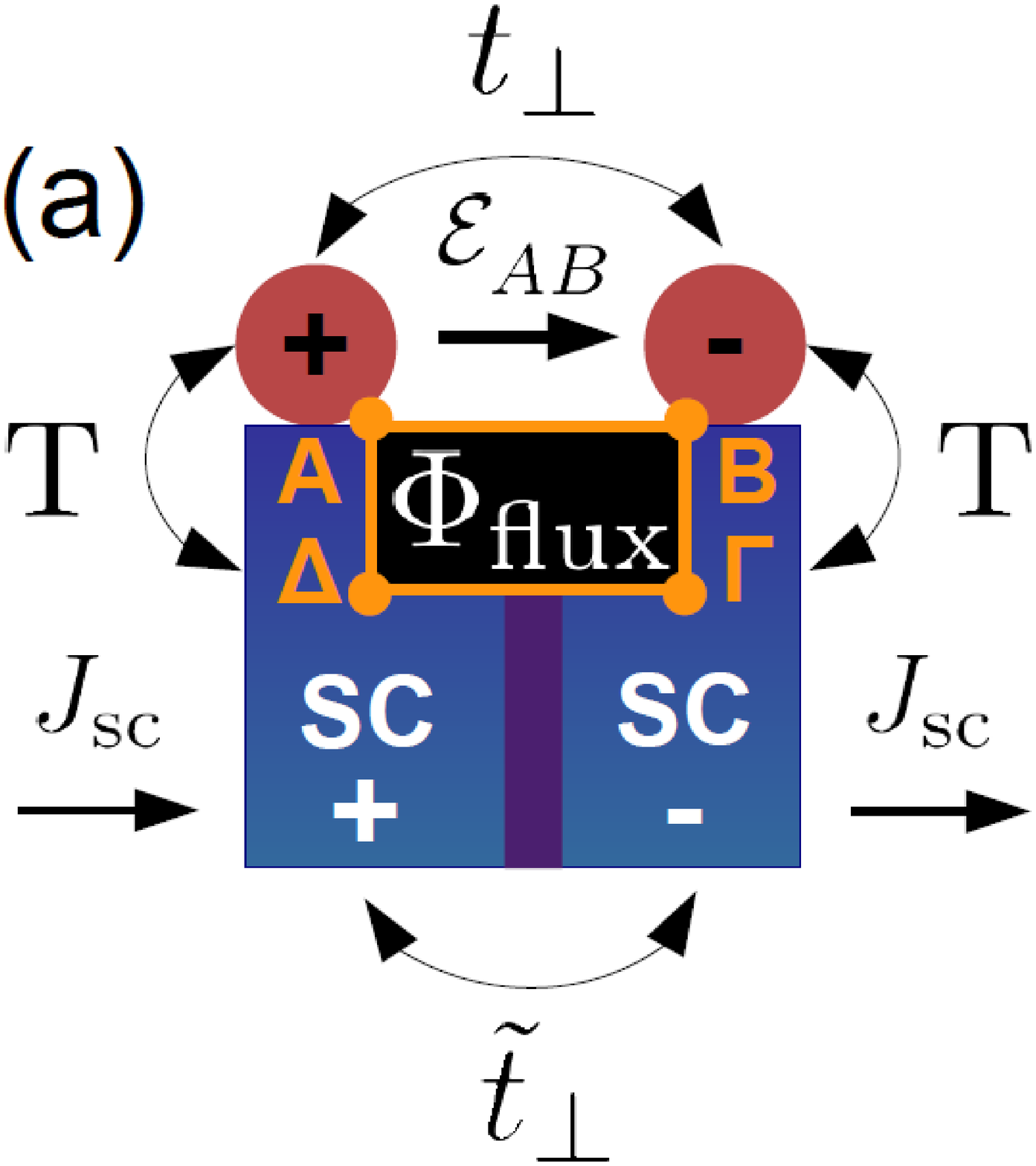}\hspace{0.1in}
\includegraphics[width=0.54\columnwidth]{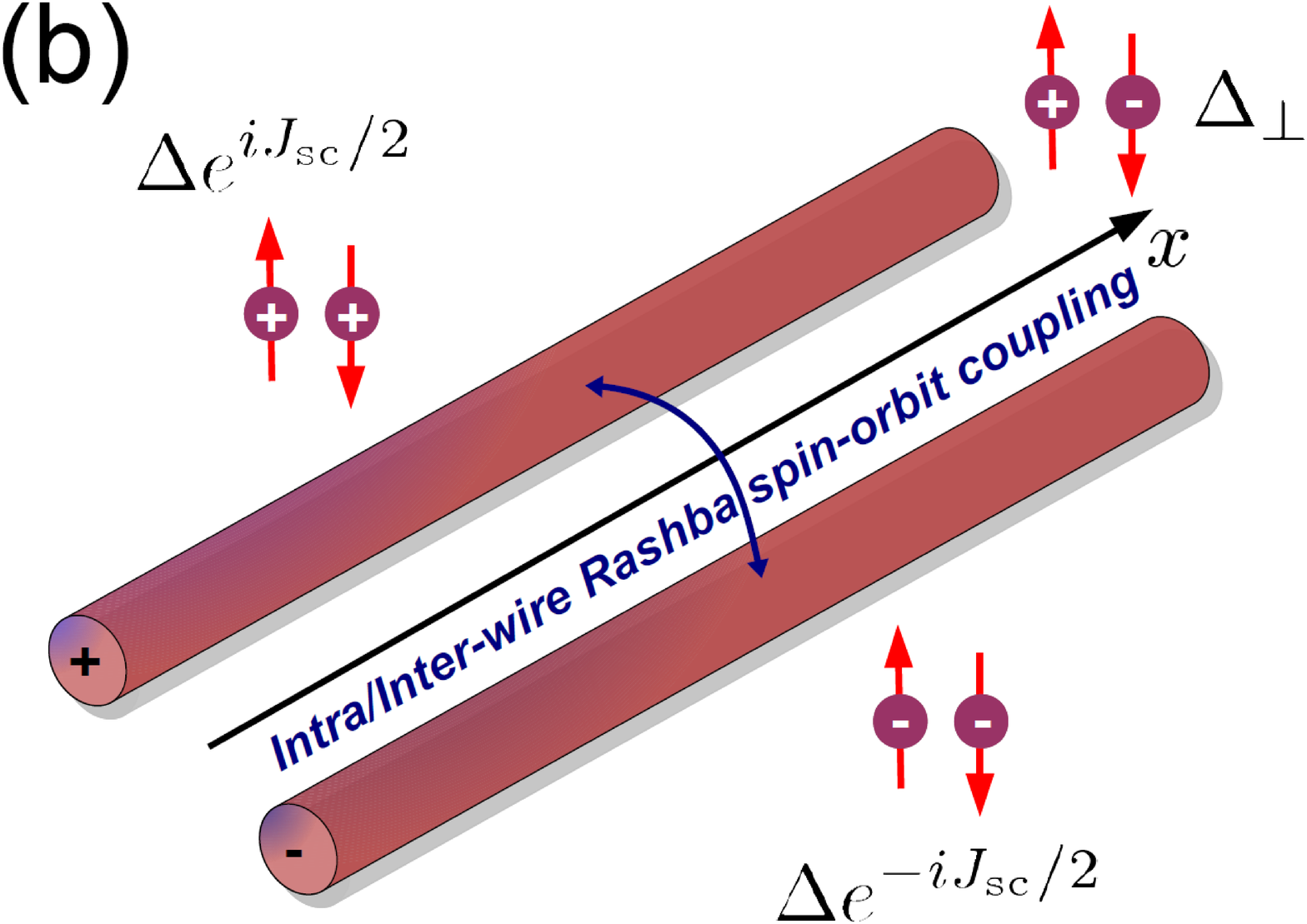}
\caption{(a) Cross-section of Fig.~\ref{fig:Setups}(b) with couplings: NW-NW ($t_{\perp}$), SC-SC ($\tilde{t}_{\perp}$), and NW-SC (${\rm T}$). Each cross-section behaves
as a superconducting quantum interference device and the total system transits to a TSC phase for a window of values for the inserted flux ($\Phi_{\rm flux}$) in the loop
(AB$\Gamma\Delta$). (b) Induced multicomponent superconduc\-ting gap on the NWs, when flux is indirectly generated by a supercurrent flow. For $J_{\rm sc}=\pi$, time-reversal
symmetry (${\cal T}$) \textit{is violated}, since the intra- and inter-NW gaps exhibit a $\pi/2$-phase locking.}
\label{fig:EffectiveWires}
\end{figure}

\noi The $\bm{\sigma}$ Pauli matrices act on spin space and the operators: $\hat{\psi}_n^{\dag}(x)=(\psi_{n\uparrow}^{\dag}(x)\,,\psi_{n\downarrow}^{\dag}(x))$ and
$\hat{c}_n^{\dag}(x)=(c_{n\uparrow}^{\dag}(x)\,,c_{n\downarrow}^{\dag}(x))$, create electrons on the NWs and the SCs, respectively. The two pa\-ral\-lel
\textit{single-channel} NWs ($n=\pm$) are placed at distance $L_y$. The terms: $\varepsilon(\hat{p}_x)=\hat{p}_x^2/2m-\mu$, $t_{\perp}$, $v$, $v_{\perp}$ de\-no\-te: kinetic
energy relative to the che\-mi\-cal potential, inter-NW hopping, intra- and inter-NW SOC. Si\-mi\-lar\-ly, $\tilde{\varepsilon}(\hat{p}_x)$ and $\tilde{t}_{\perp}$ provide the
analogous terms for the electrons of the SCs. The above Hamiltonian is obtained for a particular gauge (see App.~\ref{App:GTNW}), in which the electric field satis\-fies
${\cal E}_{AB}=\dot{\Phi}_{\rm flux}$ and ${\cal E}_{\Delta\Gamma}=0$. Here $\phi=\Phi_{\rm flux}/\Phi_0$ denotes the norma\-lized flux ($\Phi_0=h/2e$). Furthermore, I
included the spin singlet superconduc\-ting order para\-me\-ters $\tilde{\Delta} e^{niJ_{\rm sc}/2}$, with a phase dif\-fe\-ren\-ce equal to the supercurrent ($J_{\rm sc}$)
flowing through the Josephson junction. Thus only the gauge invariant quantities, $\Phi_{\rm flux}$ and $J_{\rm sc}$, appear in the Hamiltonian. A cross-section of the
heterostructure con\-sti\-tu\-tes a superconduc\-ting quantum interference device with the coupled NWs playing the role of the se\-cond weak link
(Fig.~\ref{fig:EffectiveWires}(a)), reminiscent of other experimentally realized setups \cite{CarbonNanotubeSQUID}.

\subsection{Flux and Supercurrent}\label{SubSec:SuperFlux}

Threa\-ding flux is usually rea\-lized via a solenoid or an electric field given by $\Phi_{\rm flux}=\int_{t_0}^t{\rm d}t'{\cal E}_{AB}(t')$. The electric induction can be
achieved using a capacitor (side gates) which dis\-charges in the presence of an appro\-pria\-tely attached resistive circuit, while $J_{\rm sc}=0$ is ensured. A discharging
event initia\-ted at $t_0$ yields, in the statio\-na\-ry case of interest, a flux $\Phi_{\rm flux}=\tau{\cal E}_{AB}(t_0)$, with $\tau$ the characte\-ri\-stic discharging
time. Nonetheless, flux can alternatively arise by indu\-cing a time-independent supercurrent flow equal to $J_{\rm sc}=2e\Phi_{\rm flux}/\hbar=2\pi\phi$ (see
App.~\ref{App:GTNW}), while the electric field is kept zero. Thus, the \textit{simultaneous} control of both quantities is required for engineering a TSC.

\subsection{Effective Hamiltonian}\label{SubSec:EffectiveNWs}

I proceed with integrating out the superconducting degrees of freedom following Ref.~\cite{ProxiSC}. I will consider here a non-zero flux $\Phi_{\rm flux}$, while 
$J_{\rm sc}=0$. To this end, I focus on the last two parts of the complete Hamiltonian, i.e. Eqs.~\eqref{eq:HC} and \eqref{eq:HPsiC} and transfer to the bonding and 
anti-bonding basis $\hat{c}_{b,a}=(\hat{c}_{+}\pm\hat{c}_-)/\sqrt{2}$, which yields
\bea
&&{\cal H}_{c}(x)=\sum_{n}^{a,b}\left[\hat{c}_n^{\dag}\left(\frac{\hat{p}_x^2}{2\tilde{m}}-\tilde{\mu}_n\right)\hat{c}_n\up+
\tilde{\Delta}\left(c_{n\uparrow}^{\dag}c_{n\downarrow}^{\dag}+{\rm H.c.}\right)\right],\quad\\
&&{\cal H}_{\psi c}(x)={\rm T}\left(
\hat{c}_b^{\dag}\ph\frac{\hat{\psi}_+\up+\hat{\psi}_-\up}{\sqrt{2}}+
\hat{c}_a^{\dag}\ph\frac{\hat{\psi}_+\up-\hat{\psi}_-\up}{\sqrt{2}}+{\rm H.c.}\right)\,,
\eea

\noi with $\tilde{\mu}_b=\tilde{\mu}-\tilde{t}_{\perp}$ and $\tilde{\mu}_a=\tilde{\mu}+\tilde{t}_{\perp}$. The latter difference in chemical potentials leads to different
density of states at the Fermi level, $\nu_{b,a}$. Moreover, in this new diagona\-li\-zed basis the bonding and anti-bonding fermions of the superconductor can be integrated 
out independently, exac\-tly as prescribed in \cite{ProxiSC}. In the latter works, it has been additionally shown that the corresponding proximity induced gaps, $\Delta_b$ and
$\Delta_a$, are proportional to the density of states $\nu_{b,a}$. By neglecting for the present discussion the arising renormalization effects \cite{ProxiSC}, we obtain the
pro\-xi\-mi\-ty induced pairing term (also the H.c.) on the NWs 
\bea
&&\Delta_b\frac{\psi_{+\uparrow}^{\dag}+\psi_{-\uparrow}^{\dag}}{\sqrt{2}}\frac{\psi_{+\downarrow}^{\dag}+\psi_{-\downarrow}^{\dag}}{\sqrt{2}}+
\Delta_a\frac{\psi_{+\uparrow}^{\dag}-\psi_{-\uparrow}^{\dag}}{\sqrt{2}}\frac{\psi_{+\downarrow}^{\dag}-\psi_{-\downarrow}^{\dag}}{\sqrt{2}}\no\\
&=&\Delta\sum_{n=\pm}\psi_{n\uparrow}^{\dag}\psi_{n\downarrow}^{\dag}+\Delta_{\perp}(\psi_{+\uparrow}^{\dag}\psi_{-\downarrow}^{\dag}+
\psi_{-\uparrow}^{\dag}\psi_{+\downarrow}^{\dag})
\eea

\noi with $\Delta=(\Delta_b+\Delta_a)/2$ and $\Delta_{\perp}=(\Delta_b-\Delta_a)/2$. By ta\-king into account that $\Delta_{b,a}\propto\nu_{b,a}$, we obtain the relation
$\Delta_{\perp}=\Delta(\nu_b-\nu_a)/(\nu_b+\nu_a)$. Thus, intra- and inter-NW spin singlet supercon\-duc\-ting gaps are pro\-xi\-mi\-ty induced on the NWs, similar to
Ref.~\cite{Klinovaja}. The effective Hamiltonian for the NWs, leading to a TSC \cite{KotetesClassi}, reads:
\bea
&&\widehat{{\cal H}}_{\rm TSC}(\hat{p}_x)=\varepsilon(\hat{p}_x)\tau_z+v\hat{p}_x\tau_z\sigma_y-\Delta\tau_y\sigma_y\no\\
&&+\left(t_{\perp}\tau_z\kappa_x-v_{\perp}\up\kappa_y\sigma_x\right)e^{i\pi\phi\tau_z\kappa_z}-\Delta_{\perp}\up\tau_y\kappa_x\sigma_y\,,
\label{eq:TSCHamiltonian}
\eea

\noi where I employed the spinor $\widehat{\Psi}^{\dag}(x)=(\psi_{+\uparrow}^{\dag}(x)$, $\psi_{+\downarrow}^{\dag}(x)$, $\psi_{-\uparrow}^{\dag}(x)$,
$\psi_{-\downarrow}^{\dag}(x)$, $\psi_{+\uparrow}\up(x)$, $\psi_{+\downarrow}\up(x)$, $\psi_{-\uparrow}\up(x)$, $\psi_{-\downarrow}\up(x))$. The $\bm{\tau}$ and $\bm{\kappa}$
Pauli matrices act on Nambu and NW ($\pm$) spaces. Note that if we would choose to induce the required flux via a supercurrent $J_{\rm sc}=2\pi\phi$, we would obtain the
equivalent description sketched in Fig.~\ref{fig:EffectiveWires}(b), with intra-NW gaps $\Delta e^{\pm iJ_{\rm sc}/2}$ and inter-NW gap $\Delta_{\perp}$.

\subsection{Symmetry analysis}\label{SubSec:Symmetry} 

In the following paragraphs I consider several subcases of the Hamiltonian in Eq.~\eqref{eq:TSCHamiltonian}, for constant $\phi$, with target to provide better understanding
of the particular model and reveal the necessary ingredients for engineering time-reversal symmetry (${\cal T}$) violating TSC phases.

\subsubsection{$\phi=\Delta_{\perp}=0:$  No possibility for TSC}

In order to expose the topological properties of the Hamiltonian in Eq.~\eqref{eq:TSCHamiltonian} and better understand the underlying mechanism for MFs, I first
consider $\phi=0$ and assume, only for demonstration purposes, that I can independently set $\Delta_{\perp}=0$ while $\Delta\neq0$. Under these conditions, the 
Hamiltonian in Eq.~\eqref{eq:TSCHamiltonian} reads: 
\bea
\widehat{{\cal H}}_{\rm TSC}(\hat{p}_x)&=&\varepsilon(\hat{p}_x)\tau_z+v\hat{p}_x\tau_z\sigma_y+t_{\perp}\tau_z\kappa_x-v_{\perp}\up\kappa_y\sigma_x\no\\
&-&\Delta\tau_y\sigma_y\,.
\eea

\noindent For the bulk system, i.e. $\hat{p}_x\rightarrow \hbar k$, the eigenspectrum becomes: 
\bea E(k)=\pm\sqrt{\Delta^2+\left[\varepsilon(k)\pm\sqrt{(v\hbar k\pm t_{\perp})^2+v_{\perp}^2}\right]^2}\,.\eea 

\noi By direct inspection we infer that the bulk eigenspectrum cannot support any type of gap closings, therefore no transition is possible from the topologically
tri\-vial phase to non-trivial ones, implying that MF-phases are not accessible. 

\subsubsection{$\phi=\pi$ and $\Delta_{\perp}=0:$  ${\cal T}-$invariant TSC}

This si\-tua\-tion radically changes if instead $\phi=1/2$ while $\Delta_{\perp}=0$. In this case, the Hamiltonian enjoys the symmetries: $\Theta=\kappa_x{\cal K}$ \&
$\widetilde{\Theta}=i\kappa_z\sigma_y{\cal K}$ (time-reversal), $\Xi=\tau_x{\cal K}$ \& $\widetilde{\Xi}=\tau_x\kappa_y\sigma_y{\cal K}$ (charge-conjuga\-tion) and 
$\Pi=\tau_x\kappa_x$ \& $\widetilde{\Pi}=\tau_x\kappa_z\sigma_y$ (chiral). Here ${\cal K}$ denotes complex conjugation. Note that $\Theta^2=\Xi^2=\widetilde{\Xi}^2=+I$ and
$\widetilde{\Theta}^2=-I$. Essentially $\widetilde{\Theta}$ effects ${\cal T}$, responsible for the emergence of MF Kramers pairs. The two chiral symmetries lead to a unitary
symmetry ${\cal O}\propto\Pi\widetilde{\Pi}=\kappa_y\sigma_y$, which commutes with the Hamiltonian, allowing its dia\-go\-na\-li\-za\-tion into two sub-blocks. Via the unitary
transformation ${\cal U}=(\kappa_z+\kappa_y\sigma_y)/\sqrt{2}$, I diago\-na\-li\-ze ${\cal O}$ and obtain the two resulting $\kappa=\pm$ sub-blocks: 
\bea
\widehat{{\cal H}}_{\kappa}^{{\cal U}}(\hat{p}_x)&=&\varepsilon(\hat{p}_x)\tau_z+v\hat{p}_x\tau_z\sigma_y+\kappa
t_{\perp}\sigma_y-v_{\perp}\up\tau_z\sigma_z\no\\
&-&\Delta\tau_y\sigma_y\,.\label{eq:MFKramers}
\eea

Each block ($\kappa=\pm$) describes a single channel Rashba NW in the presence of a supercon\-duc\-ting gap $\Delta$ and a block dependent Zeeman field
$\bm{{\cal B}}_{\kappa}=(0,\kappa t_{\perp},-v_{\perp})$, with parallel and perpendicular components to the SOC orientation. \textit{In this gauge}, the supercurrent (or flux)
converts the inter-NW Rashba SOC into a Zeeman term which is oriented perpendicular to the intra-NW SOC $v\hat{p}_x\tau_z\sigma_y$ \cite{KotetesClassi}, as in strictly 1d NW
mo\-dels \cite{NW}. Therefore $v_{\perp}$ is here a \textit{prerequisite} for TSC (this holds also for the general $\phi\neq\pi$ and $\Delta_{\perp}\neq0$ case as discussed
in App.~\ref{App:interNWSOC}). Both blocks belong to symmetry class ${\rm D}$, with the charge conjugation symmetry $\Xi=\tau_x{\cal K}$. If the value of $t_{\perp}$ is such,
so that the energy spectrum is fully gapped, each subsystem harbors a single MF per edge when $\sqrt{t_{\perp}^2+v_{\perp}^2}>\sqrt{\Delta^2+\mu^2}$, associated with the bulk
energy spectrum closing at the inversion symmetric point $k=0$. Due to the underlying ${\cal T}$, the two subsystems transit to the topologically non-trivial phase for the
same parameter values, and thus the resulting two MFs per edge constitute a Kramers pair similar to Refs.~\cite{Doublewire}. 

However, in contrast to Refs.~\cite{Doublewire} which focused on ${\cal T}$-preserving TSCs, here I go beyond and study ${\cal T}$-violating TSCs which become accessible
after the inclusion of either $\Delta_{\perp}$ or deviations of $\phi$ from the value $1/2$. 

\subsubsection{$\phi\neq\pi$ and $\Delta_{\perp}=0:$  Splitting of the MF Kramers pair}\label{par:noTSC}

Introducing deviations of the flux from the value $\phi=1/2$, yields the ${\rm BDI}$ symmetry class with symmetries: $\{\Xi,\Theta,\Pi\}$. It is instructive to study the
consequence of {\textit small} deviations $\lambda$, with $\phi=1/2+\lambda/\pi$, on the pre\-existing MF Kramers pair related to $k=0$. The Majorana wavefunctions can
be retrieved by setting $\hat{p}_x=0$ in Eq.~\eqref{eq:MFKramers}, i.e. $-\mu\tau_z+t_{\perp}\kappa_z\sigma_y-v_{\perp}\up\tau_z\sigma_z-\Delta\tau_y\sigma_y=0$. By
introducing $\tan\delta=\mu/\Delta$ and $\tan\beta=t_{\perp}/v_{\perp}$, the MF related eigenvectors have the form
${\rm Exp}[-i(\delta\tau_x\sigma_y+\beta\kappa_z\tau_z\sigma_x)/2](1/\sqrt{2})\widehat{{\cal X}}$ with $\widehat{{\cal X}}$ (note that \textit{here} the eigenvectors are
written for convenience in $\kappa\otimes\tau\otimes\sigma$ space):
\bea
\left(\begin{array}{cccccccc}0&1&1&0&0&0&0&0\end{array}\right)^T\,,
\left(\begin{array}{cccccccc}0&0&0&0&0&1&1&0\end{array}\right)^T\,,\qquad\label{eq:noTSC1}\\
\left(\begin{array}{cccccccc}1&0&0&1&0&0&0&0\end{array}\right)^T\,,
\left(\begin{array}{cccccccc}0&0&0&0&1&0&0&1\end{array}\right)^T\,.\qquad\label{eq:noTSC2}
\eea

\noindent The eigenspectrum of the Hamiltonian confined in the above subspace for $\hat{p}_x=0$, reads:
\bea
\pm
\sqrt{\left(\sqrt{t_{\perp}^2+v_{\perp}^2}-\sqrt{\mu^2+\Delta^2}\right)^2+\lambda^2\left(t_{\perp}^2+v_{\perp}^2\right)}\,.
\eea

\noindent For $\lambda=0$ the eigenergies touch when $\sqrt{t_{\perp}^2+v_{\perp}^2}=\sqrt{\mu^2+\Delta^2}$, providing two MF modes. If we switch on $\lambda$, no touching 
can occur and the possibility for zero modes vanishes. Remarkably, one finds that for an arbitrary phase $\phi$, the spectrum has always a minimum value, i.e.
$|E(k)|\geq\Delta|\cos(\pi\phi)|$ $\forall k$, implying that there can be no gap closing for $\phi\neq1/2$, as the MFs Kramers pair splits. 

\subsubsection{$\phi=\pi$ and $\Delta_{\perp}\neq0:$  ${\cal T}-$violating TSC}

The situation changes when a non-zero $\Delta_{\perp}$ is added to Eq.~\eqref{eq:TSCHamiltonian}, not allowing the block diagonalization of the Hamiltonian as in
Eq.~\eqref{eq:MFKramers}. Instead it yields a single block Hamiltonian residing once again in BDI class with symmetries: $\{\Xi,\Theta,\Pi\}$. By considering a small
$\Delta_{\perp}$, and residing on the analysis of Sec.~\ref{par:noTSC} (Eqs.~\eqref{eq:noTSC1},\eqref{eq:noTSC2}), I find that the modification on the energy spectrum of 
the preexisting MF Kramers pair of $k=0$, now becomes:
\bea
&&\pm\sqrt{\left(\sqrt{t_{\perp}^2+v_{\perp}^2}-\sqrt{\mu^2+\Delta^2}\right)^2+\Delta_{\perp}^2\sin^2\beta\sin^2\delta}\no\\
&&\pm\Delta_{\perp}\cos\beta\cos\delta\,.
\eea

\noi By inspecting the equation above, it is straightforward to discern that the addition of a small $\Delta_{\perp}$ splits the MF Kramers pair but nevertheless allows for 
a single tou\-ching and thus a single MF mode. Evenmore for large $\Delta_{\perp}$ the system can have the possibility of supporting multiple MFs due to the presence of chiral
symmetry in BDI class. In the particular case, one can have up to two MFs per edge, since each MF Kramers pair per edge can still survive in the presence of $\Delta_{\perp}$,
with each one of the previous MF Kramers pair partners now originating from the inversion-symmetry connected points $\pm k_*\neq0$. 

\subsubsection{$\phi\neq\pi$ and $\Delta_{\perp}\neq0:$  ${\cal T}-$violating TSC}

In the most general case discussed here, MFs are possible due to the presence of $\Delta_{\perp}$ which counterbalances the effect of the deviations of the flux from the
value $\phi=1/2$. Thus, the presence of $\Delta_{\perp}$ is vital for preserving TSC away from $\phi=1/2$, and can even sustain a pair of MFs per edge in spite of the lifted
Kramers dege\-ne\-ra\-cy. In fact, for the parameters considered in the present analysis, $\Delta_{\perp}$ allows the emergence of MFs in the window $0.43<\phi\leq1/2$.

\subsection{TSC mechanism}\label{SubSec:TSCmech}

According to the present proposal, crucial ingredient for engineering a ${\cal T}-$violating TSC phase is the combined presence of a supercurrent (or $\Phi_{\rm flux}$) 
and the inter-NW SC. On one hand, the supercurrent converts the inter-NW SOC into a Zeeman term, while on the other, $\Delta_{\perp}$ gua\-ran\-tees that TSC can survive 
away from the critical pa\-ra\-me\-ter space point $\phi=1/2$. Notably ${\cal T}$ is broken even for $\phi=1/2$, which constitutes a situation equi\-va\-lent to inducing a
supercurrent flow that reali\-zes a $\pi$-junction, depicted in Fig.~\ref{fig:EffectiveWires}(b). From this point of view, ${\cal T}-$violation can be understood as
$\pi/2$-locking of the multicomponent superconducting gap, since the intra-NW gaps become ima\-gi\-na\-ry ($\pm i\Delta$) and the inter-NW gap remains real ($\Delta_{\perp}$).
Thus the multicomponent superconducting gap violates ${\cal T}$, offering a unique mecha\-nism for obtaining single-MF phases \textit{without} a Zeeman field. Finally, note
that a supercurrent flow has been also recently proposed in Ref. \cite{AFMShiba} as an indi\-spen\-sable ingre\-dient for engineering a TSC, while other works
\cite{supercurrents,supercurrents2} have highlighted its utility for either tailoring \cite{supercurrents} or mapping out \cite{supercurrents2} MF-phases. 

\subsection{Topological phase diagram}\label{SubSec:TPD} 

By virtue of chiral symmetry, I block off-diagonalize the Hamiltonian of Eq.~\eqref{eq:TSCHamiltonian} via the transformation $(\tau_z+\tau_x\kappa_x)/\sqrt{2}$. I obtain the
$k$-space Hamiltonian 
\bea
\widehat{{\cal H}}'(k)= \tau_+\up\hat{A}(k)+\tau_-\up\hat{A}^{\dag}(k)\,,\eea

\noi with the off-diagonal block projectors $\tau_{\pm}=(\tau_x\pm i\tau_y)/2$ and 
\bea
\hat{A}(k)=\varepsilon(k)\kappa_x+v\hbar k\kappa_x\sigma_y+\left(v_{\perp}\sigma_x-it_{\perp}\kappa_z\right)\sin\left(\pi\phi\right)\no\\
+\left(t_{\perp}+iv_{\perp}\kappa_z\sigma_x\right)\cos\left(\pi\phi\right)-i(\Delta+\Delta_{\perp}\kappa_x)\sigma_y\,.\quad\label{eq:Ak}
\eea

\noi The relevant $\mathbb{Z}$ topological invariant, ${\cal N}$, is defined as the winding number of $D(k)\equiv{\rm Det}[\hat{A}(k)]$ \cite{Tewari and Sau}. By employing 
the unit vector $\hat{\bm{g}}(k)=\left(0,D_{\Im}(k)/D(k),D_{\Re}(k)/D(k)\right)$ (see App.~\ref{App:TopoInv}), ${\cal N}$ is defined as \cite{AFMShiba}
\bea
{\cal N}=\frac{1}{2\pi}\int dk\ph\left(\hat{\bm{g}}(k)\times\frac{\partial\hat{\bm{g}}(k)}{\partial k}\right)_x\,.\label{eq:TopInv}
\eea

\begin{figure}[t]
\includegraphics[width=1.0\columnwidth]{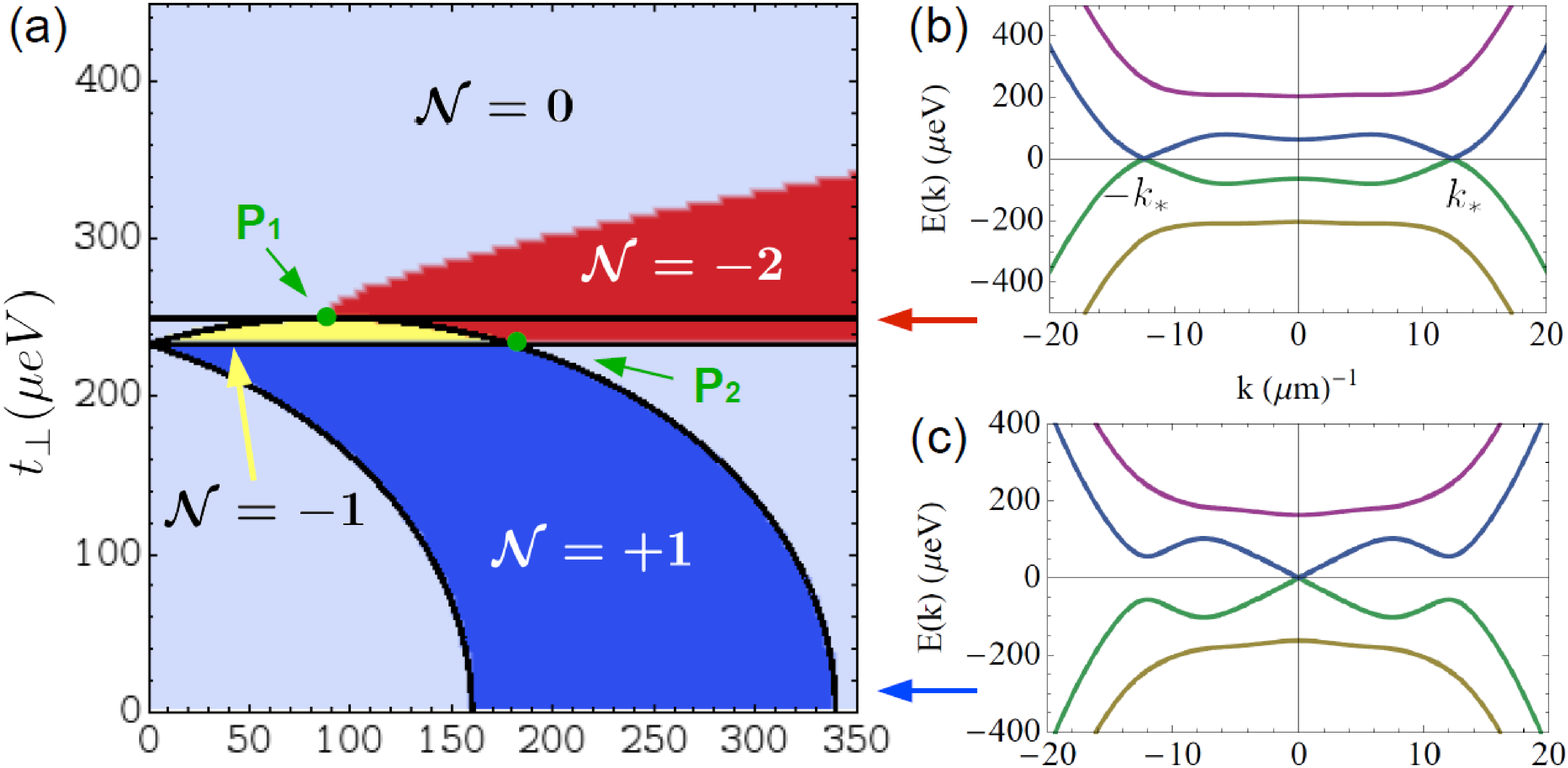}\vspace{0.1in}
\includegraphics[width=0.49\columnwidth]{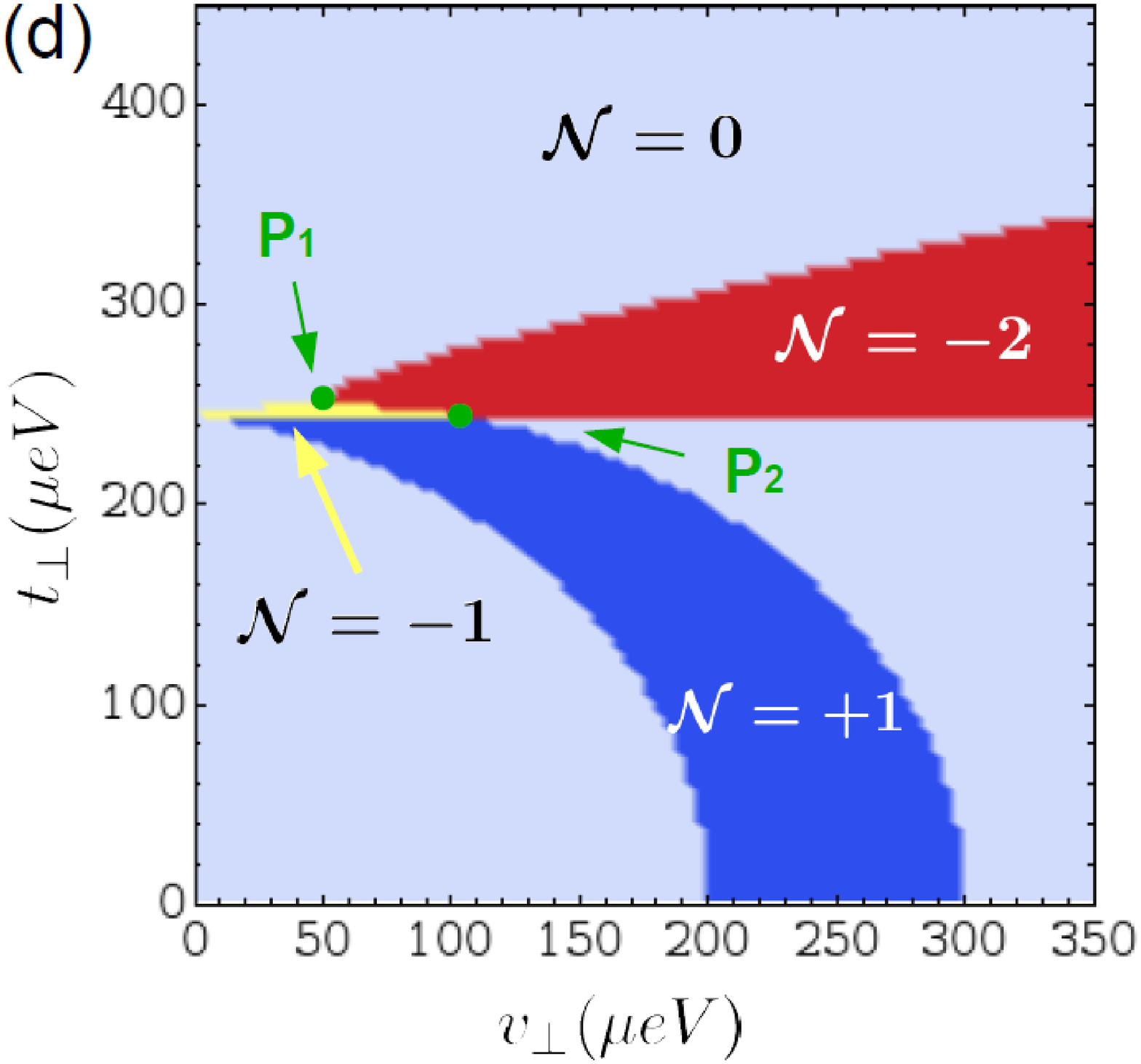}\hspace{0.0in}
\includegraphics[width=0.49\columnwidth]{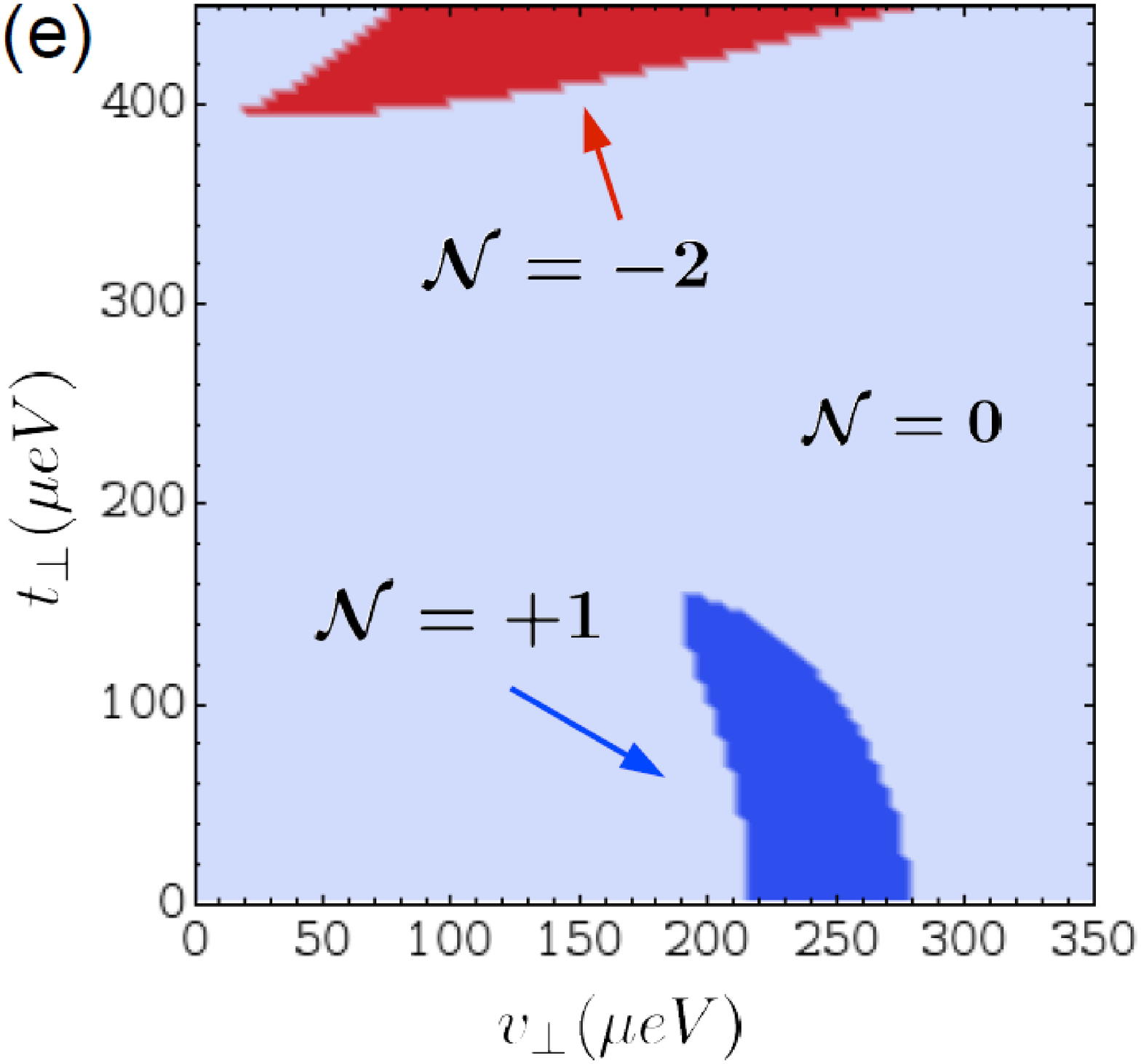}
\centering
\caption{Topological phase diagrams for InSb and InAs ($\mu=0$, $\Delta=250{\rm \mu eV}$). (a) $\Delta_{\perp}=90{\rm \mu eV}$ and $\phi=1/2$ ($J_{\rm sc}=\pi$). Four distinct
phases appear: ${\cal N}=0,\pm 1,2$, with $|{\cal N}|$ the number of MFs per edge. The 1MF-phases (blue \& yellow) are enclosed by the critical lines
$\Delta^2=t_{\perp}^2+(\Delta_{\perp}\pm v_{\perp})^2$, defined by the bulk gap closings at $k=0$, as in (c) for $(v_{\perp},t_{\perp})=(300,136){\rm \mu eV}$.
The ${\cal N}=-2$ phase is protected by chiral symmetry and arises from gap closings at $\pm k_*$, as in (b) for $(v_{\perp},t_{\perp})=(300,233){\rm \mu eV}$. The ${\cal
N}=+1$ and ${\cal N}=-2$ phases, overlap, yielding the 1MF-phase with ${\cal N}=-1$. The latter appears between the parallel lines:
$t_{\perp}=\sqrt{\Delta^2-\Delta_{\perp}^2}$ and $t_{\perp}\simeq\Delta$. Two quantum tricritical points emerge: $P_1\simeq(\Delta_{\perp},\Delta)$ and
$P_2=(2\Delta_{\perp},\sqrt{\Delta^2-\Delta_{\perp}^2})$, where the 0,1 and 2 MF-phases meet. (d) $\Delta_{\perp} =50{\rm \mu eV}$ and $\phi=1/2$. Results 
similar to (a), but with the window for the 1MF-phase suppressed as it depends on $\Delta_{\perp}$. For $\Delta_{\perp}=0$, the 1MF-phase disappears, since Kramers degeneracy
is restored and only MF pairs are allowed. (e) $\Delta_{\perp}=50{\rm \mu eV}$ and $\phi=0.45$. Further away from the $\pi$-junction, the critical points $P_{1,2}$ vanish.
}
\label{fig:PhaseDiagram}
\end{figure}

In Fig.~\ref{fig:PhaseDiagram}, I present a series of $v_{\perp}-t_{\perp}$ topological phase diagrams (for further details see App.~\ref{App:TopoInv}), obtained for $\mu=0$
and representative values for InSb NWs with pro\-xi\-mi\-ty induced supercon\-ducti\-vi\-ty, i.e. $v\hbar=0.20{\rm eV\AA}$, $m=0.015m_e$ and $\Delta=250{\rm \mu eV}$
\cite{Mourik}. In Fig.~\ref{fig:PhaseDiagram}(a) $\phi=1/2$ and $\Delta_{\perp}=90{\rm \mu eV}$. Here we find phases with 0, 1 or 2 MFs per edge. The single MF phase is
bounded by the lines $\Delta^2=t_{\perp}^2+(\Delta_{\perp}\pm v_{\perp})^2$, which are given by the gap closing condition for the inversion-symmetric wave-vector $k=0$
(Fig.~\ref{fig:PhaseDiagram}(c)). On the other hand, the phase with two MFs is asso\-cia\-ted with gap clo\-sings at two non-inversion-symmetric wave-vectors, $\pm k_*$, which
are however connected by inversion (Fig.~\ref{fig:PhaseDiagram}(b)). The phase with 2 MFs is topo\-lo\-gi\-cal\-ly protected as long as chiral symmetry $\Pi=\tau_x\kappa_x$
persists \cite{Tewari and Sau}. This symmetry could be vio\-lated by a mismatch in the intra-NW superconduc\-ting gaps (or chemical potentials), which would enter in
Eq.~\eqref{eq:TSCHamiltonian} with a term $\sim\tau_y\kappa_z\sigma_y$ ($\sim\tau_z\kappa_z$). 

For $(\hbar k)^2/2m\gg v\hbar k$, one finds that the gap closings at $\pm k_*$ occur at parts of the parallel lines $t_{\perp}=\sqrt{\Delta^2-\Delta_{\perp}^2}$ and
$t_{\perp}\simeq\Delta$, shown in Fig.~\ref{fig:PhaseDiagram}(a). The phases with ${\cal N}=+1$ and ${\cal N}=-2$ overlap, leading to the phase with ${\cal N}=-1$. As a
result, two quantum tri\-cri\-ti\-cal points emerge (see also \cite{CriticalPoints}), where the phases with 0, 1 and 2 MFs meet. The latter appear at
$P_1\simeq(\Delta_{\perp},\Delta)$ and $P_2=(2\Delta_{\perp},\sqrt{\Delta^2-\Delta_{\perp}^2})$. The phase diagram of Fig.~\ref{fig:PhaseDiagram}(d(e)) was retrieved for
$\Delta_{\perp}=50{\rm \mu eV}$ and $\phi=1/2(0.45)$. The tricritical points exist only near $\phi=1/2$. Evenmore, away from this value the TSC region becomes suppressed. If
$t_{\perp}\simeq0$, the window for a single MF-phase is given by $\Delta-\Delta_{\perp}<v_{\perp}<\Delta+\Delta_{\perp}$. Thus maximizing $\Delta_{\perp}$ enhances the
robustness of the TSC phase.

\subsection{Predictions for experiments}\label{SubSec:ExpNWs} 

In reality the two NWs have a finite diameter $d\sim110{\rm nm}$ \cite{Mourik,NWcrosses}. If they are in contact, we can assume that $t_{\perp}$ and $v_{\perp}$
are given by the expectation value of the kinetic energy and Rashba SOC component related to hoppings along the $y$-direction. Therefore, in this case I can set
$t_{\perp}=<\hat{p}_y^2/2m>\sim \hbar^2/(2mL_y^2)$ and $v_{\perp}=<v\hat{p}_y>\sim v\hbar/L_y$, where for a rough estimation I assumed $<\hat{p}_y^{s}>\sim(\hbar/L_y)^s$ 
with $s=1,2$. Under these assumptions, I obtain that for $\phi=1/2$ and $\Delta_{\perp}=90 (50){\rm \mu eV}$, the 1MF-phase can be rea\-lized if $105{\rm nm}<L_y<150{\rm nm}$
($110{\rm nm}<L_y<130{\rm nm}$). Setting instead $v\hbar=0.15{\rm eV\AA}$ and $m=0.024m_e$, addresses the InAs case. By resca\-ling $k$, one finds that the topological phase
diagrams for InAs almost coincide with those of Fig.~\ref{fig:PhaseDiagram}. For $\phi=1/2$ and $\Delta_{\perp}=90(50){\rm \mu eV}$, the 1MF-phase appears in InAs NWs for
$80{\rm nm}<L_y<115{\rm nm}$ ($85{\rm nm}<L_y<100{\rm nm}$). Thus the 1MF-phase is accessible \textit{only} when the NWs are placed \textit{in contact to each other}
($L_y\simeq d$). Otherwise, $t_{\perp}$ and $v_{\perp}$ are associated with electron tunneling and thus become much weaker, as they are given by the exponentially decaying
overlap of the NW wavefunctions. Finally, the 2MF-phase appears experimentally inaccessible for the \textit{particular} setup, since it would require for the NWs to be closer
than their dia\-meter ($L_y<d$).

\section{Two-channel Rashba semiconducting film on top of a Jo\-se\-phson jun\-ction}\label{Sec:2Cs}

\begin{figure}[b]
\centering
\includegraphics[width=0.6\columnwidth]{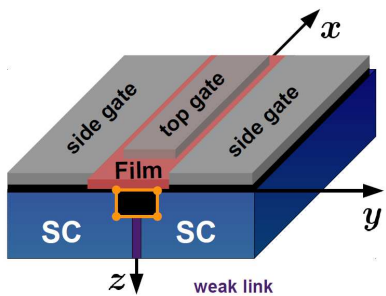}
\caption{Full view of a Rashba semiconducting film, on top of a Josephson junction. The interface is interrupted by an insulating loop enclosed by the yellow lines. Threading
magnetic flux through the loop, for instance by effecting an electric field using the side gates, leads to Majorana bound states extended along the $y$ axis and localized at
the edges along the $x$ axis. The top gate can be employed for tuning the chemical potential of the semiconducting film and thus controlling the topological phase diagram.}
\label{fig:Figure_4}
\end{figure}

In the present section I focus on a TSC device consi\-sting of a Rashba semiconducting film (Sec.~\ref{SubSec:Film}). Its geometry is quasi-1d due to hard-wall confinement
along the $y$ axis as in Fig.~\ref{fig:Figure_4}. I consider an effective model constrained to the two lowest confinement channels (Sec.~\ref{SubSec:H2C}). The two-channel
model can be mapped to the coupled NW system (Sec.~\ref{SubSec:EffectiveNWs}), and thus, the symmetry and topologi\-cal studies of the previous paragraphs
(Secs.~\ref{SubSec:Symmetry}-\ref{SubSec:TPD}) apply also here. Based on the latter, I predict the design details of such devices for harboring either one or two MFs per edge
(Sec.~\ref{SubSec:Exp2C}).

\subsection{Effective Hamiltonian}\label{SubSec:Film}

The effective Hamiltonian for a Rashba semiconduc\-ting film with proximity induced conventional superconductivity, has the form:
\bea
{\cal H}&=&\int
d\bm{r}\ph\hat{\psi}^{\dag}(\bm{r})\left[\frac{\hat{\bm{\pi}}^2}{2m}-\mu(t,y)+v\left(\hat{\bm{\pi}}\times\bm{\sigma}\right)\cdot\hat{\bm{z}}\right]\hat{\psi}(\bm{r})\no\\
&+&\int d\bm{r}\ph\Delta(y)\left[\psi_{\uparrow}^{\dag}(\bm{r})\psi_{\downarrow}^{\dag}(\bm{r})+\psi_{\downarrow}\up(\bm{r})\psi_{\uparrow}\up(\bm{r})\right]\,,
\label{eq:Hamiltonian}
\eea

\noi where I introduced the gauge invariant canonical momentum 
\bea
\hat{\bm{\pi}}=\hat{\bm{p}}+e\bm{A}+\hbar\bm{\nabla}\varphi/2\,.\eea

\noi Here $\bm{A}=A_y\hat{\bm{y}}$ defines the vector potential in the film (equivalent to $\Phi_{AB}$ in App.~\ref{App:GTNW}), while $\varphi=\varphi(t,y)$ is the
phase of the bulk SC. Moreover, I introduced the gauge invariant chemical (scalar) potential
\bea
\mu(t,y)=\mu-\hbar\dot{\varphi}/2-U_{\rm conf}(y)+eV_{\rm sg}(t,y)\,,
\eea

\noi consisting of a chemical potential $\mu$, a superconduc\-ting phase contribution, a confining potential $U_{\rm conf}(y)$ and the electrostatic side gate potential 
$V_{\rm sg}(t,y)$. Here I consider a confining potential $U_{\rm conf}(y)=+\infty$ for $y<0$ and $y>L_y$, and zero otherwise. Its form allows employing the confinement
channel wavefunctions $\left<y|n\right>=\sqrt{2/L_y}\sin(n\pi y/L_y)$ with $n=1,2,\ldots$ and $\epsilon_n\equiv\left<n|\hat{p}_y^2/2m|n\right>=(\hbar \pi n)^2/2mL_y^2$. For
$v_{\perp}=0$ they constitute exact eigenstates of the effective Hamiltonian, while for $v_{\perp}\neq0$ they can be employed as a complete basis for furnishing a
representation of the latter.

The present effective Hamiltonian has been retrieved by integrating out the superconducting degrees of freedom as in Sec.~\ref{SubSec:EffectiveNWs} and according to
Ref.~\cite{ProxiSC}. Markedly, the superconducting pro\-xi\-mity effect is partially blocked, yielding an induced gap $\Delta(y)$ which is zero for $y\in
[(L_y-b)/2,(L_y+b)/2]$ and equal to $\bar{\Delta}$ otherwise ($b\equiv$(AB); see Fig.~\ref{fig:Setups}(c)). The particular spatial profile is \textit{required} for
re\-trie\-ving a ${\cal T}$-violating multicomponent gap, similar to $\Delta$ and $\Delta_{\perp}$ encountered in the NW case. Instead, if the proximity is not blocked,
$\Delta(y)=\Delta$ and no superconducting gap equivalent to $\Delta_{\perp}$ can appear. 

By assuming that the vector potential across the interface is zero (equivalent to $\Phi_{A\Delta}=\Phi_{B\Gamma}=0$ of App.~\ref{App:GTNW}) and that the spatial dependence
of the superconducting phase is of the form $\bm{\nabla}\varphi=\partial_y\varphi\hat{\bm{y}}$ we obtain the relation
\bea
\int_{0}^{L_y}\left(A_y+\frac{\hbar}{2e}\frac{\partial\varphi}{\partial y}\right)dy=-\Phi_{\rm flux}\,.\label{eq:Flux}
\eea

\noi Similarly to the NW case, I set $V_{\rm sg}(t,y)=\hbar\dot{\varphi}/2e$, and obtain 
\bea
\int_0^{L_y}{\cal E}_ydy=\dot{\Phi}_{\rm flux}\,.
\eea

\subsection{Projection onto the two-lowest confinement channels}\label{SubSec:H2C}

Since the confinement channel wavefunctions do not constitute eigenstates of Eq.~\eqref{eq:Hamiltonian}, I will consider an approximate model Hamiltonian constructed by 
the two lowest confinement channels. The validity of the latter Hamiltonian strongly depends on the position of the chemical potential and is expected to break down for 
large $L_y$, as in this case the energy level differences become small. Here, I intend to place the chemical potential symmetrically between $\epsilon_1$ and $\epsilon_2$, 
i.e. $\mu=5\epsilon_1/2$, which is achie\-vable with appropriate gating. This value allows focusing on the two-lowest channels, while it additionally constitutes a sweet spot
\cite{multiband}, for which charge fluctuations are suppressed.   

Care has to be taken, so that the approximate Hamiltonian follows the same gauge transformation rules as the parent Hamiltonian of Eq.~\eqref{eq:Hamiltonian}. In fact, the
mi\-ni\-mal coupling scheme $\hat{\bm{p}}\rightarrow\hat{\bm{p}}+e\bm{A}$ must be properly modified. For this reason, I first project the Hamiltonian of
Eq.~\eqref{eq:Hamiltonian} onto the lowest confinement channels, for $V_{\rm sg}=A_y=\varphi=0$. This yields
\bea
\widehat{{\cal H}}_{\rm film}(\hat{p}_x)=
\left(\frac{\hat{p}_x^2}{2m}-\mu+\frac{\epsilon_2+\epsilon_1}{2}\right)\tau_z+v\hat{p}_x\tau_z\sigma_y\no\\
+\frac{\epsilon_2-\epsilon_1}{2}\ph\tau_z\kappa_z-\frac{8v\hbar}{3L_y}\kappa_y\sigma_x\qquad\qquad\phd\ph\no\\
-\frac{\Delta_2+\Delta_1}{2}\ph\tau_y\sigma_y-\frac{\Delta_2-\Delta_1}{2}
\ph\tau_y\kappa_z\sigma_y\,,\label{eq:H2cLowest}
\eea

\noi with the $\bm{\kappa}$ Pauli matrices acting on the channel subspace $n=\{2,1\}$ and $\Delta_n=\int_0^{L_y}dy\left<y|n\right>^2\Delta(y)$ given by
\bea
\Delta_2&=&\bar{\Delta}\left[1-b/L_y+\sin(\pi b/L_y)\cos(\pi b/L_y)/\pi\right]\,,\\
\Delta_1&=&\bar{\Delta}\left[1-b/L_y-\sin(\pi b/L_y)/\pi\right]\,.
\eea

In order to introduce the gauge potentials, I will first retrieve the expression for the polarization operator in this basis. The polarization operator reads $P_y=-ey$ and in
this basis has the representation $P_y=16eL_y\tau_z\kappa_x/9\pi^2$. In the presence of a homogeneous time-dependent electric field, ${\cal E}_y$, the Hamiltonian acquires an
additional $-P_y{\cal E}_y$ term. In the latter case, one can infer the coupling of the two-channel system with the electrostatic potential and the superconducting phase,
which reads, $P_y\partial_y(V_{\rm sg}-\hbar\dot\varphi/2e)$. Here I assume that the gradients of the electrostatic potential and superconducting phase are constants. To
retrieve the coupling to the vector potential, $A_y$, we have to first obtain the expression for the paramagnetic current operator $J_y=\dot{P}_y$. The latter time derivative
can be retrieved using the Heisenberg equation of motion for the polarization operator calculated using the non-superconducting, and therefore gauge invariant, part of the
Hamiltonian in Eq.~\eqref{eq:H2cLowest}. Thus, we have 
\bea 
J_y&=&\frac{i}{\hbar}\left[\left(\frac{\hat{p}_x^2}{2m}-\mu+\frac{\epsilon_2+\epsilon_1}{2}\right)\tau_z+v\hat{p}_x\tau_z\sigma_y\right.\no\\
&&\left.+\frac{\epsilon_2-\epsilon_1}{2}
\ph\tau_z\kappa_z-\frac{8v\hbar}{3L_y}\kappa_y\sigma_x,\frac{16eL_y}{9\pi^2}\tau_z\kappa_x\right]\no\\
&=&-\frac{2}{\hbar}\frac{16eL_y}{9\pi^2}\left(\frac{\epsilon_2-\epsilon_1}{2}\ph\kappa_y+\frac{8v\hbar}{3L_y}\tau_z\kappa_z\sigma_x\right)\,.
\eea

\noi The current above provides the linear correction to the Hamiltonian with respect to $A_y+\hbar\partial_y\varphi/2e$. Consequently, the interchannel terms become modified 
in the following manner
\bea
&&\frac{\epsilon_2-\epsilon_1}{2}\ph\tau_z\kappa_z-\frac{8v\hbar}{3L_y}\kappa_y\sigma_x\rightarrow\no\\
&&\frac{\epsilon_2-\epsilon_1}{2}\ph\tau_z\kappa_z-\frac{8v\hbar}{3L_y}\kappa_y\sigma_x-J_y\big(A_y+\hbar\partial_y\varphi/2e\big)\no\\
&=&
\left(\frac{\epsilon_2-\epsilon_1}{2}\ph\tau_z\kappa_z-\frac{8v\hbar}{3L_y}
\kappa_y\sigma_x\right)\times\no\\
&&\left[1+i\frac{2}{\hbar}\frac{16eL_y}{9\pi^2}\left(A_y+\frac{\hbar}{2e}\frac{\partial\varphi}{\partial y}\right)\tau_z\kappa_x\right]\,.
\eea

\noi The above linear term is useful for calculating the linear response to the external fields, e.g. conductivities, but can not yield the desired gauge transformation
properties. To serve the latter purpose it has to get exponen\-tia\-ted, i.e.  
\bea
&&\frac{\epsilon_2-\epsilon_1}{2}\ph\tau_z\kappa_z-\frac{8v\hbar}{3L_y}\kappa_y\sigma_x\rightarrow\\
&&\left(\frac{\epsilon_2-\epsilon_1}{2}\ph\tau_z\kappa_z-\frac{8v\hbar}{3L_y}
\kappa_y\sigma_x\right){\rm Exp}\left(-i\frac{32}{9\pi^2}\frac{e\Phi_{\rm flux}}{\hbar}\tau_z\kappa_x\right)\,,\no
\eea

\noi where I additionally made use of Eq.~\eqref{eq:Flux} by conside\-ring that $A_y$ is spatially homogeneous. If we now set the total electrostatic potential to zero, i.e.
$V_{\rm sg}-\hbar\dot{\varphi}/2e=0$, Eq.~\eqref{eq:Flux} additionally provides $L_y{\cal E}_y=\dot{\Phi}_{\rm flux}\Rightarrow\Phi_{\rm flux}=L_y\int_{t_0}^tdt'{\cal
E}_y(t')$. Therefore, the normalized flux in the particular case reads $\phi=-(32/9\pi^2)\Phi_{\rm flux}/\Phi_0$. We imme\-diately notice the difference compared to the
NW case, in which $\phi=\Phi_{\rm flux}/\Phi_0$. The projection onto the lowest two confinement channels yields an effective flux piercing the loop AB${\rm \Gamma\Delta}$,
equal to $-32\Phi_{\rm flux}/9\pi^2\simeq-0.36\Phi_{\rm flux}$. 

Under the aforementioned conditions and after effec\-ting the unitary transformation $(\kappa_z+\kappa_x)/\sqrt{2}$, the Hamiltonian becomes 
\bea
&&\widehat{{\cal H}}_{\rm film}'(\hat{p}_x)=
\left[\frac{\hat{p}_x^2}{2m}-\left(\mu-\frac{5\epsilon_1}{2}\right)\right]\tau_z+v\hat{p}_x\tau_z\sigma_y-\Delta_c\tau_y\sigma_y\no\\
&&+\left(\frac{3\epsilon_1}{2}\tau_z\kappa_x+\frac{8v\hbar}{3L_y}\kappa_y\sigma_x\right)e^{i\pi\phi\tau_z\kappa_z}-\frac{\delta\Delta}{2}\ph\tau_y\kappa_x\sigma_y\,,
\quad\label{eq:H2cLowestFinal}
\eea

\noi where I introduced the average $\Delta_c=\left(\Delta_2+\Delta_1\right)/2$ and difference $\delta\Delta=\Delta_2-\Delta_1$ of the superconducting gaps, given by  
\bea
&&\Delta_c=\bar{\Delta}\left[1-b/L_y-\sin(\pi b/L_y)\sin^2\left(\pi b/2L_y\right)/\pi\right]\,,\quad\\
&&\delta\Delta/2=\bar{\Delta}\sin(\pi b/L_y)\cos^2\left(\pi b/2L_y\right)/\pi\,.
\eea

\noi Therefore, this Hamiltonian can be mapped to the one of Eq.~\eqref{eq:TSCHamiltonian}, with the parameters of Eq.~\eqref{eq:H2cLowestFinal} having the correspondence: 
\bea
\mu\rightarrow\mu-\frac{5\epsilon_1}{2}\,,\quad
t_{\perp}\rightarrow\frac{3\epsilon_1}{2}\,,\quad
v_{\perp}\rightarrow-\frac{8v\hbar}{3L_y}\,,\no\qquad\quad\\
\Delta\rightarrow\Delta_c\,,\phd 
\Delta_{\perp}\rightarrow\frac{\delta\Delta}{2}\,,\phd
\phi\rightarrow-\frac{32L_y}{9\pi^2\Phi_0}\int_{t_0}^tdt'{\cal E}_y(t')\,.\quad
\eea

\noi As observed in the analysis of the NW case, we can maximize the TSC window via maximizing $\Delta_{\perp}$ which here corresponds to $\delta\Delta$. For the
two-channel model, this occurs for $b/L_y=1/3$, yielding $\Delta_c\simeq3\bar{\Delta}/5$ and $\delta\Delta/2\simeq\bar{\Delta}/5$. In fact, the latter optimal values will
be assumed for the discussion below.

\subsection{Predictions for experiments}\label{SubSec:Exp2C} 

The most prominent rea\-li\-za\-tion of this setup is based on already existing 2DEG devices \cite{2DEGs}. I assume that $\phi=1/2$, $\mu=5\epsilon_1/2$, while I set 
$\bar{\Delta}=400{\rm \mu eV}$, which implies $\Delta_c=240{\rm \mu eV}$ and $\delta\Delta/2=80{\rm \mu eV}$. Under these conditions, the 1MF-phase is stabilized for 
InAs when $315{\rm nm}<L_y<390{\rm nm}$. On the other hand, for InSb the 1MF-phase appears for $400{\rm nm}<L_y<500{\rm nm}$. In stark contrast to the double-NW case, here the
2MF-phase \textit{becomes} ex\-peri\-men\-tally \textit{accessible}, ap\-pro\-xi\-ma\-te\-ly when: $295{\rm nm}<L_y<310{\rm nm}$ for InAs and $370{\rm nm}<L_y<395{\rm nm}$ for
InSb.

\section{Three-channel Rashba semiconducting nanowire on top of a Jo\-se\-phson jun\-ction}\label{Sec:3Cs}

\begin{figure}[t]
\centering
\includegraphics[width=0.6\columnwidth]{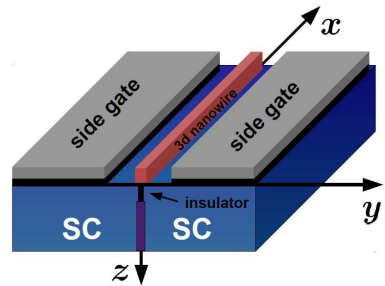}
\caption{Full view of a three-dimensional Rashba semiconducting nanowire with square crossection ($yz$ plane) on top of a Josephson junction. The interface is interrupted by
an insulating part (depicted with black) which partially blocks the proximity effect and renders the proximity induced superconductivity inhomogeneous along the $y$ axis.
Using the side gates, one can induce an electric field in the semiconductor which in the steady state can support Majorana bound states extended along the $y$ axis and
localized at the edges along the $x$ axis.}
\label{fig:Figure_5}
\end{figure}

In this section I consider a hybrid device shown in Fig.~\ref{fig:Figure_5} in which a three-dimensional Rashba semiconducting NW, confined in the $yz$ plane, lies on top of a
Josephson junction (Sec.~\ref{SubSubSec:HWire}). For the particular SOC type, the low ener\-gy description of the NW requires the consideration of the lowest three-channels
(Sec.~\ref{SubSec:H3C}), on which I focus after having integrated out the superconducting degrees of freedom. According to my analysis, only two of the above three channels
become coupled and can in principle exhibit TSC phases, similarly to Sec.~\ref{Sec:2Cs}. Using the previous analysis I conclude that the experimental rea\-li\-za\-tion
(Sec.~\ref{SubSec:ExpWire}) of artificial TSC phases in such systems require a quite stong proximity induced gap, which appears not accessible, at least for the moment, using
\textit{conventional} SCs, e.g., Al, Nb or Pb.

\subsection{Effective Hamiltonian}\label{SubSubSec:HWire}

In this paragraph I present the effective Hamiltonian for a hybrid device involving a three-dimensional Rashba semiconducting wire. For simplicity, I will here consider a wire
with square cross-section $L_y=L_z\equiv d$. The starting point is the Hamiltonian
\bea
&&{\cal H}=\int d\bm{r}\ph\hat{\psi}^{\dag}(\bm{r})\left[\frac{\hat{\bm{\pi}}^2}{2m}-\mu(t,y,z)+
v\left(\hat{\bm{\pi}}\times\bm{\sigma}\right)\cdot\hat{\bm{z}}\right]\hat{\psi}(\bm{r})\no\\
&&+\int d\bm{r}\ph\Delta(y)\left[\psi_{\uparrow}^{\dag}(\bm{r})\psi_{\downarrow}^{\dag}(\bm{r})+\psi_{\downarrow}\up(\bm{r})\psi_{\uparrow}\up(\bm{r})\right]\,,
\label{eq:H3C}
\eea

\noi with $\hat{\bm{\pi}}=\hat{\bm{p}}+e\bm{A}+\hbar\bm{\nabla}\varphi/2$ and $\mu(y,z)=\mu-\hbar\dot{\varphi}/2-U_{\rm conf}(y,z)+eV_{\rm sg}(t,y)$. Here I assumed that the
superconducting gap varies \textit{only} along the $y$ axis, due to the blocked proximity effect by an insulating layer as in the case of the semiconducting film. The
conside\-ra\-tion of an infinite well confining potential $U_{\rm conf}(y,z)=0$ for $\{y,z\}\in\{[0,d],[0,d]\}$ and $+\infty$ otherwise, allows us to introduce the confinement
channel wavefunctions $\left<y,z|n,s\right>=(2/d)\sin(n\pi y/d)\sin(s\pi z/d)$ with $n,s=1,2,\ldots$ and
$\epsilon_{n,s}\equiv\left<n,s|\left(\hat{p}_y^2+\hat{p}_z^2\right)/2m|n,s\right>=(\hbar \pi)^2(n^2+s^2)/2md^2$. The latter constitute eigenstates of the above Hamiltonian
for $v_{\perp}=0$ while they can be used as a complete basis set for $v_{\perp}\neq0$. As in the previous section, I will pursue a low-energy description also here. The
energetically lowest level is $(1,1)$ (correspon\-ding to the set of quantum numbers $(n,s)$) while the two energetically higher, $(2,1)$ and $(1,2)$, are de\-ge\-ne\-ra\-te.
Thus three confinement channels have to be taken into account.

\subsection{Projection onto the three-lowest confinement channels}\label{SubSec:H3C}

As in Sec.~\ref{SubSec:H2C}, the projection on the three-lowest channels will be performed first for $A_y=\varphi=V_{\rm sg}=0$. 
\bea
&&\widehat{{\cal H}}_{\rm wire}(\hat{p}_x)=
\left(\frac{\hat{p}_x^2}{2m}-\mu+\epsilon_1+\frac{\epsilon_{2}+\epsilon_{1}}{2}\right)\tau_z+v\hat{p}_x\tau_z\sigma_y\no\\
&&+
\frac{\epsilon_{2}-\epsilon_{1}}{2}\ph\tau_z\left(\begin{array}{ccc}1&0&0\\0&1&0\\0&0&-1\end{array}\right)
-\frac{8v\hbar}{3L_y}\left(\begin{array}{ccc}0&0&0\\0&0&-i\\0&i&0\end{array}\right)\sigma_x\no\\
&&-\frac{\Delta_{2}+\Delta_{1}}{2}\ph\tau_y\sigma_y
-\frac{\Delta_{2}-\Delta_{1}}{2}\ph\tau_y\left(\begin{array}{ccc}-1&0&0\\0&1&0\\0&0&-1\end{array}\right)\sigma_y\,,\quad\label{eq:H3cLowest}
\eea

\noi with the basis $\{(1,2),(2,1),(1,1)\}$. Note that I used $\epsilon_{1,2}$ and $\Delta_{1,2}$ defined in Sec.~\ref{SubSec:Film}, with $L_y=d$. As seen from the above, in
the absence of gauge potentials the channel $(1,2)$ decouples from the other two. In order to infer if this situation persists for finite gauge potentials, I calculate the
polarization operators $P_y=-ey$ and $P_z=-ez$ in this basis. I retrieve
\bea
P_y=\frac{16ed}{9\pi^2}\left(\begin{array}{ccc}0&0&0\\0&0&1\\0&1&0\end{array}\right)\,,\phd
P_z=\frac{16ed}{9\pi^2}\left(\begin{array}{ccc}0&0&1\\0&0&0\\1&0&0\end{array}\right)\,.\quad
\eea

For the situation considered in this manuscript, the electric field ${\cal E}_z$ in the semiconducting wire is zero. Thus for only ${\cal E}_y$ finite, the channel $(1,2)$
decouples even in the presence of gauge potentials and the Hamiltonian for the two remaining coupled channels, $(2,1)$ and $(1,1)$, is identical to the two-channel film case
of Sec.~\ref{SubSec:Film}. Note that the channel $(1,2)$ can not support MFs. If the proximity induced gap becomes $z$ dependent, then it is possible for the $(1,1)$ and
$(1,2)$ channels to couple. Nonetheless, even in the latter case the presence of the $(1,2)$ channel does not change the qualitative topological characteristics and only
brings some quantitative modifications to the phase diagram. In order for this channel to become topologically relevant, a different type of SOC has to considered, which
includes also the $\hat{p}_z$ momentum. As a matter of fact, only inter-channel coupling induced by SOC which will be converted into an effective Zeeman term, can lead to new
topological properties due to the addition of the $(1,2)$ channel.

\subsection{Predictions for experiments}\label{SubSec:ExpWire}

By focusing on the two coupled channels, I infer the boundaries for phases with a single MF per edge using the relations $\Delta^2=t_{\perp}^2+(\Delta_{\perp}\pm
v_{\perp})^2$, as in Sec.~\ref{SubSec:TPD}. For $b=L_y/3$ and a che\-mi\-cal potential symmetrically placed inbetween the levels $(2,1)$ and $(1,1)$, as in
Sec.~\ref{SubSec:Film}, I find that the phase with a single MF per edge is realized for a proximity induced gap: $5.2{\rm meV}<\bar{\Delta}<5.8{\rm meV}$. Thus it seems
\textit{currently} not feasible to engineer a TSC with a single InSb wire, via the proposed mechanism. The reason is the large energy splitting $\sim3{\rm meV}$ of the two
coupled channels, compared to the recently expe\-ri\-mentally achieved induced superconducting gap using conventional SCs $\sim 0.6{\rm meV}$ \cite{NWcrosses}.

\section{Conclusions}\label{Sec:Conclusions}

To summarize, I proposed a new class of MF platforms relying on low-dimensional semiconductors, which allow \textit{replacing} the Zeeman field with supercurrents or electric
fields. Double-nanowire setups, which can support 1MF-phases when the nanowires are parallel and in contact to each other, appear experimentally accessible \cite{NWcrosses}.
In contrast, devices based on a three-channel three-dimensional nanowire demand a large proximity induced superconducting gap and can become experimentally feasible if high-Tc
superconductors, such as the ones based on Fe \cite{Hosono}, can be employed. On the other hand, versatile InAs 2DEG devices in proximity to a Josephson junction which have
been already fabricated and manipulated three decades ago, constitute ideal candidates for rea\-li\-zing the two-channel implementation. They can exhibit an interplay of
phases with 1 or 2 MFs per edge, depen\-ding on the width of the device. Optimization purposes require a supercurrent value of $J_{\rm sc}=\pi$, which can be achieved by
connec\-ting the Josephson junction to a large superconducting ring threaded by flux. Alternatively, an electric field $\tau{\cal E}_{y}(t_0)L_y=9\pi^2\Phi_0/64$ can be
applied across the film. For $\tau\sim 1\mu s({\rm ns})$ and $L_y\sim 400{\rm nm}$, a weak field $\sim 70{\rm \mu V(mV)/cm}$ is required, opening new perspectives for
\textit{all-electrical} control on MF devices.\\

\begin{acknowledgments}

I am glad to thank A. Shnirman, G. Sch\"{o}n, A. Geresdi, R. Aguado, V. Mourik, E. Prada, A. Heimes, D. Pikulin, M. Wimmer, J. Schmalian, P.-Q. Jin, D. Mendler and T.
Schmidt for valua\-ble discus\-sions and suggestions which significantly helped me to improve and complete this work.

\end{acknowledgments}

\begin{widetext}

\appendix

\section{Gauge transformation properties of the double nanowire Hamiltonian: Flux \& Supercurrent}\label{App:GTNW}

For clarity, I introduce in this section an extended and more general form of the hybrid device Hamiltonian of Sec.~\ref{SubSec:HD2NW}:
\bea
{\cal H}_{\psi}(x)&=&\sum_{n=\pm}\hat{\psi}_n^{\dag}\left[\varepsilon(\hat{p}_x)+v\hat{p}_x\sigma_y-ne\frac{V_{AB}}{2}\right]\hat{\psi}_n\up+
\hat{\psi}_+^{\dag}\left(t_{\perp}+iv_{\perp}\sigma_x\right)e^{i\frac{e}{\hbar}\Phi_{AB}}\hat{\psi}_-\up+
\hat{\psi}_-^{\dag}\left(t_{\perp}-iv_{\perp}\sigma_x\right)e^{-i\frac{e}{\hbar}\Phi_{AB}}\hat{\psi}_+\up\,,\quad\\
{\cal H}_{c}(x)&=&\sum_{n=\pm}\left\{\hat{c}_n^{\dag}\left[\tilde{\varepsilon}(\hat{p}_x)-ne\frac{V_{\Delta\Gamma}}{2}\right]\hat{c }_n\up+
\tilde{\Delta}\left(e^{i\varphi_n}c_{n\uparrow}^{\dag}c_{n\downarrow}^{\dag}+e^{-i\varphi_n} c_{n\downarrow}\up c_{n\uparrow}\up\right)\right\}
+\tilde{t}_{\perp}\left(e^{i\frac{e}{\hbar}\Phi_{\Delta\Gamma}}\hat{c}_+^{\dag}\hat{c}_-\up
+e^{-i\frac{e}{\hbar}\Phi_{\Delta\Gamma}}\hat{c}_-^{\dag}\hat{c}_+\up\right),\quad\phd\ph\ph\\
{\cal H}_{\psi c}(x)&=&{\rm T}
\left(e^{i\frac{e}{\hbar}\Phi_{A\Delta}}\hat{\psi}_+^{\dag}\hat{c}_+\up+e^{i\frac{e}{\hbar}\Phi_{B\Gamma}}\hat{\psi}_-^{\dag}\hat{c}_-\up+
e^{-i\frac{e}{\hbar}\Phi_{A\Delta}}\hat{c}_+^{\dag}\hat{\psi}_+\up+e^{-i\frac{e}{\hbar}\Phi_{B\Gamma}}\hat{c}_-^{\dag}\hat{\psi}_-\up\right)\,.
\eea

\noi To ensure full generality I have added appropriate voltage drops ($V_{ab}$) and Peierls'-phases ($\Phi_{ab}$) for all the pairwise coupled ele\-ments $(a,b)$ of the
hybrid device. For compactness I have suppressed the $x$ dependence of the field operators.

For illustrating the connection between the flux piercing the loop AB${\rm \Gamma\Delta}$ and the supercurrent flow, I perform the following gauge
transformation $\hat{\psi}_n(x)=e^{nie\chi/2\hbar}\hat{\psi}_n'(x)$ and $\hat{c}_n(x)=e^{nie\alpha/2\hbar}\hat{c}_n'(x)$, with $\chi$ and $\alpha$ independent of $x$. In the
new gauge: $V_{AB}'=V_{AB}\up-\dot{\chi}$, $\Phi_{AB}'=\Phi_{AB}\up-\chi$, $V_{\Delta\Gamma}'=V_{\Delta\Gamma}\up-\dot{\alpha}$,
$\Phi_{\Delta\Gamma}'=\Phi_{\Delta\Gamma}\up-\alpha$, $\Phi_{A\Delta}'=\Phi_{A\Delta}\up-(\chi-\alpha)/2$, $\Phi_{B\Gamma}'=\Phi_{B\Gamma}\up+(\chi-\alpha)/2$ and
$\varphi_n'=\varphi_n\up-ne\alpha/\hbar$, where $\dot{f}$ denotes the time derivative of $f$. To this end, I demand:
$V_{AB}'=V_{\Delta\Gamma}'=\Phi_{\Delta\Gamma}'=\Phi_{A\Delta}'=\Phi_{B\Gamma}'=0$, which imposes: ${\cal E}_{AB}=\dot{\Phi}_{\rm flux}$, ${\cal E}_{\Delta\Gamma}=0$,
$\Phi_{AB}'=\Phi_{AB}\up-\chi\equiv-\Phi_{\rm flux}\up$, $\varphi_n'=\varphi_n-ne\alpha/\hbar\equiv\varphi_n-ne\Phi_{\Delta\Gamma}/\hbar$, with the electric field ${\cal
E}_{ab}=V_{ab}-\dot{\Phi}_{ab}$. This additionally provides $\delta\varphi'=\delta\varphi-2e\Phi_{\Delta\Gamma}/\hbar=J_{\rm sc}$, i.e., equal to the supercurrent $J_{\rm sc}$
flowing through the junction. Thus only $\Phi_{AB}'\equiv-\Phi_{\rm flux}$ and $\delta\varphi'\equiv J_{\rm sc}$, persist in the gauged Hamiltonian, considered also in
Sec.~\ref{SubSec:HD2NW} of the manuscript. For the rest, I will consider that $\Phi_{\rm flux}$ and $J_{\rm sc}$ are time-independent. In the steady state, their connection
can be demonstrated by perfor\-ming the additional gauge transformation $\chi\rightarrow\chi-\Phi_{\rm flux}$ and $\alpha\rightarrow\alpha-\Phi_{\rm flux}$, yiel\-ding
$\Phi_{\rm flux}\rightarrow 0$ and $J_{\rm sc}\rightarrow J_{\rm sc}+2e\Phi_{\rm flux}/\hbar$. Therefore, threa\-ding flux $\Phi_{\rm flux}$ is also equivalent to inducing 
a supercurrent flow $J_{\rm sc}=2e\Phi_{\rm flux}/\hbar$. 

\section{Necessary requirement of inter-NW SOC for engineering a TSC}\label{App:interNWSOC}

If we set $v_{\perp}=0$ in the Hamiltonian of the main text given by Eq.~\eqref{eq:TSCHamiltonian}, we find that the latter commutes with $\sigma_y$ and can be diagonalized
into two blocks which belong to the AIII symmetry class with the chiral symmetry operator $\tau_x\kappa_x$. By introducing the eigenvectors of $\sigma_y$, labelled by
$\sigma=\pm1$, Eq.~\eqref{eq:Ak} of the main text can be also block diagonalized with $\hat{A}_{\sigma}(k)=[\varepsilon(k)+\sigma v\hbar
k-i\sigma\Delta_{\perp}]\kappa_x-it_{\perp}\sin\left(\pi\phi\right)\kappa_z+t_{\perp}\cos\left(\pi\phi\right)-i\sigma\Delta$. Via calculating the winding number of the
unit-vectors $\hat{\bm{g}}_{\sigma}(k)=(0,{\rm Det}_{\Im}[\hat{A}_{\sigma}(k)]/|{\rm Det}[\hat{A}_{\sigma}(k)]|,{\rm Det}_{\Re}[\hat{A}_{\sigma}(k)]/|{\rm
Det}[\hat{A}_{\sigma}(k)]|)$, as in Eq.~\eqref{eq:TopInv}, we find that the system always lies in the topologically trivial phase if $v_{\perp}=0$. 

\section{Topological invariant and bulk gap closings}\label{App:TopoInv}

For $\mu=0$ and $\phi=1/2$ one obtains the expressions for the $\bm{g}(k)$ vector of Sec.~\ref{SubSec:TPD}:
\bea
g_y(k)&=&4\Delta_{\perp}v(\hbar k)\left[t_{\perp}^2+\Delta_{\perp}^2-v_{\perp}^2-\Delta^2-v^2(\hbar k)^2+\frac{1}{(2m)^2}(\hbar k)^4\right]\,,\\
g_z(k)&=&\left[t_{\perp}^2+(\Delta_{\perp}+v_{\perp})^2-\Delta^2\right]\left[t_{\perp}^2+(\Delta_{\perp}-v_{\perp})^2-\Delta^2\right]+
2v^2\left(v_{\perp}^2-3\Delta_{\perp}^2-t_{\perp}^2+\Delta^2\right)(\hbar k)^2\no\\&+&
\left[v^4+\frac{2}{(2m)^2}\left(\Delta^2+\Delta_{\perp}^2-v_{\perp}^2-t_{\perp}^2\right)\right](\hbar k)^4-2(v/2m)^2(\hbar k)^6+\frac{1}{(2m)^4}(\hbar k)^8\,.
\eea

\noindent The zeroes of $\bm{g}(k)$ provide the topological phase boundaries and the bandstructure points where the related gap closing occurs. The gap closings at $k=0$ give
rise to a single MF while the gap closings at $(\hbar k_*)^2/2m=mv^2\pm\sqrt{(mv^2)^2+\Delta^2+v_{\perp}^2-\Delta_{\perp}^2-t_{\perp}^2}$ provide two MFs, which are
protected by chiral symmetry instead of Kramers degeneracy. The phase with a single MF is enclosed within the area defined by the lines
$t_{\perp}^2+(\Delta_{\perp}\pm v_{\perp})^2=\Delta^2$ (dark blue and yellow phase in Fig.~3 of the manuscript), while the phase with two MFs is enclosed within the area
defined by the lines $t_{\perp}=\sqrt{\Delta^2-\Delta_{\perp}^2}$, $t_{\perp}=\sqrt{\Delta^2-(\Delta_{\perp}-v_{\perp})^2}$ and
$t_{\perp}=\sqrt{\Delta^2+2mv^2\sqrt{v_{\perp}^2-\Delta_{\perp}^2}}$ (red phase in Fig.~3 of the manuscript). The phase with the two MFs becomes topologically trivial when
chiral symmetry is broken and the system transits to class D, which can happen if an asymmetry between the intra-wire superconducting gaps or chemical potentials is
introduced.


\end{widetext}


\begin{thebibliography}{00}

\bibitem{TQC}A. Yu. Kitaev, Annals Phys. \bt{303}, 2 (2003); C. Nayak, S. H. Simon, A. Stern, M. Freedman, and S. Das Sarma, Rev. Mod. Phys. \bt{80}, 1083 (2008); 
N. Read and D. Green, Phys. Rev. B \bt{61}, 10267 (2000); A. Yu. Kitaev, Phys.-Usp. \textbf{44}, 131 (2001); D. A. Ivanov, Phys. Rev. Lett. \bt{86}, 268 (2001); L. Fu and C.
L. Kane, \textit{ibid.} \textbf{100}, 096407 (2008); J. Alicea, Y. Oreg, G. Refael, F. von Oppen, and M. P. A. Fisher, Nature Physics \bt{7} 412 (2011); J. Alicea, Rep. Prog.
Phys. \bt{75}, 076501 (2012); C. W. J. Beenakker, Annu. Rev. Con. Mat. Phys. \bt{4}, 113 (2013). 

\bibitem{Semi}J. D. Sau, R. M. Lutchyn, S. Tewari, and S. Das Sarma, Phys. Rev. Lett. \textbf{104}, 040502 (2010); J. Alicea, Phys. Rev. B \textbf{81} 125318 (2010).

\bibitem{NW}R. M. Lutchyn, J. D. Sau and S. Das Sarma, Phys. Rev. Lett. \textbf{105}, 077001 (2010); Y. Oreg, G. Refael and F. von Oppen, Phys. Rev. Lett. \textbf{105},
177002 (2010).

\bibitem{multiband}R. M. Lutchyn, T. Stanescu, S. Das Sarma, Phys. Rev. Lett. \bt{106}, 127001 (2011). 

\bibitem{Tewari and Sau}S. Tewari and J. D. Sau, Phys. Rev. Lett. \bt{109}, 150408 (2012).

\bibitem{Doublewire}S. Deng, L. Viola, and G. Ortiz, Phys. Rev. Lett. \bt{108}, 036803 (2012); S. Nakosai, Y. Tanaka, and N. Nagaosa, Phys. Rev. Lett. \bt{108}, 147003 (2012);
A. Keselman, L. Fu, A. Stern, and E. Berg, Phys. Rev. Lett. \bt{111}, 116402 (2013); E. Gaidamauskas, J. Paaske, and K. Flensberg, Phys. Rev. Lett. \bt{122}, 126402 (2014); A.
Haim, A. Keselman, E. Berg, and Y. Oreg, Phys. Rev. B \bt{89}, 220504(R) (2014).

\bibitem{Klinovaja}J. Klinovaja and D. Loss, Phys. Rev. B \bt{90}, 045118 (2014).

\bibitem{Frustaglia}A. A. Reynoso, and D. Frustaglia, Phys. Rev. B \bt{87}, 115420 (2013).

\bibitem{KotetesClassi}P. Kotetes, New J. Phys. \textbf{15}, 105027 (2013).

\bibitem{Mourik} V. Mourik, K. Zuo, S. M. Frolov, S. R. Plissard, E. P. A. M. Bakkers, and L. P. Kouwenhoven, Science \bt{336}, 1003 (2012).

\bibitem{MFexperiments}M. T. Deng, C. L. Yu, G. Y. Huang, M. Larsson, P. Caroff, and H. Q. Xu, Nano Lett. \bt{12}, 6414 (2012); L. P. Rokhinson, Xinyu Liu, and J. K.
Furdyna, \textit{ibid.} \bt{8}, 795 (2012); A. Das, Y. Ronen, Y. Most, Y. Oreg, M. Heiblum, and Hadas Shtrikman, \textit{ibid.} \bt{8}, 887 (2012). 

\bibitem{MFexperiments2}E. J. H. Lee, X. Jiang, R. Aguado, G. Katsaros, C. M. Lieber, and S. De Franceschi, Phys. Rev. Lett. \bt{109}, 186802 (2012); A. D. K. Finck, D. J. Van
Harlingen, P. K. Mohseni, K. Jung, and X. Li, \textit{ibid.} \bt{110}, 126406 (2013); H. O. H. Churchill, V. Fatemi, K. Grove-Rasmussen, M. T. Deng, P. Caroff, H. Q. Xu, and
C. M. Marcus, Phys. Rev. B \bt{87}, 241401(R) (2013); E. J. H. Lee, X. Jiang, M. Houzet, R. Aguado, C. M. Lieber, and S. De Franceschi, Nat. Nanotechnol. \bt{9}, 79 (2014).

\bibitem{Flensberg}M. Leijnse and K. Flensberg, Phys. Rev. Lett. \bt{107}, 210502 (2011).

\bibitem{2DEGs}H. Takayanagi and T. Kawakami, Phys. Rev. Lett. \bt{54}, 2449 (1985); M. Thomas, H.-R. Blank, Ki C. Wong, H. Kroemer, and E. Hu, Phys. Rev. B \bt{58}, 11676
(1998).

\bibitem{BDI}R. Wakatsuki, M. Ezawa, Y. Tanaka, and N. Nagaosa, Phys. Rev. B \bt{90}, 014505 (2014); H.-Y. Hui, P. M. R. Brydon, J. D. Sau, S. Tewari, and S. Das Sarma,
Sci. Rep. \bt{5}, 8880 (2015).

\bibitem{CarbonNanotubeSQUID}J.-P. Cleuziou, W. Wernsdorfer, V. Bouchiat, T. Ondar\c{c}uhu, and M. Monthioux, Nat. Nanotechnol. \bt{1}, 53 (2006). 

\bibitem{ProxiSC} T. D. Stanescu, J. D. Sau, R. M. Lutchyn, and S. Das Sarma, Phys. Rev. B \bt{81}, 241310(R) (2010); A. C. Potter and P. A. Lee,  Phys. Rev. B \bt{83}, 
184520 (2011).

\bibitem{AFMShiba}A. Heimes, P. Kotetes, and G. Sch\"on, Phys. Rev. B \bt{90}, 060507(R) (2014). 

\bibitem{supercurrents} B. Seradjeh and E. Grosfeld Phys. Rev. B \textbf{83}, 174521 (2011); A. Romito, J. Alicea, G. Refael and F. von Oppen, Phys. Rev. B \textbf{85},
020502 (2012); X.-J. Liu and A. M. Lobos, Phys. Rev. B \textbf{87}, 060504 (2013); J. R\"ontynen and T. Ojanen, Phys. Rev. B \bt{90}, 180503(R) (2014).

\bibitem{supercurrents2}P. San-Jose, E. Prada and R. Aguado, Phys. Rev. Lett. \bt{112}, 137001 (2014). 

\bibitem{CriticalPoints}R. M. Lutchyn and M. P. A. Fisher, Phys. Rev. B \bt{84}, 214528 (2011).  

\bibitem{NWcrosses} S. R. Plissard, I. van Weperen, D. Car, M. A. Verheijen, G. W. G. Immink, J. Kammhuber, L. J. Cornelissen, D. B. Szombati, A. Geresdi, S. M. Frolov, 
L. P. Kouwenhoven, and E. P. A. M. Bakkers, Nat. Nanotechnol. \bt{8}, 859 (2013).

\bibitem{Hosono} Y. Kamihara, T. Watanabe, M. Hirano, and H. Hosono, J. Am. Chem. Soc. \bt{130} 3296 (2008).


\end{thebibliography}
\end{document}